\documentclass[aps,nofootinbib, showpacs]{revtex4-1}
\usepackage[CJKbookmarks, pdftex, bookmarksnumbered, bookmarksopen, colorlinks, citecolor=blue, linkcolor=blue]{hyperref}
\usepackage[english]{babel}
\usepackage{amstext,amsmath,amssymb,amsfonts,bbm}
\usepackage[latin1]{inputenc}
\usepackage{graphicx}
\usepackage{epstopdf}
\usepackage{amsthm}
\usepackage{tocvsec2}
\usepackage{enumerate}
\usepackage{subfigure}
\usepackage{bm}
\usepackage{color}
\usepackage{tikz}
\usepackage[squaren,Gray]{SIunits}

\newcommand{\va}{\scriptscriptstyle}

\newcommand{\be}{\nopagebreak[3]\begin{equation}}
\newcommand{\ee}{\end{equation}}

\newcommand{\bee}{\nopagebreak[3]\begin{equation*}}
\newcommand{\eee}{\end{equation*}}

\newcommand{\ba}{\nopagebreak[3]\begin{eqnarray}}
\newcommand{\ea}{\end{eqnarray}}
\DeclareFontFamily{U}{rsfs}{}         
\DeclareFontShape{U}{rsfs}{m}{n}{<5> rsfs5 <6><7> rsfs7          %
  <8><9><10><10.95><12><14.4><17.28><20.74><24.88> rsfs10}{}     %
\DeclareMathAlphabet{\mathfs}{U}{rsfs}{m}{n}                     %
\newcommand{\mfs}[1]{\mathfs {#1}}                               %
\newcommand{\n}{{\nonumber}}

\newcommand{\sH}{{\mfs H}}

\newcommand{\sN}{{\mfs N}}
\newcommand{\sM}{{\mfs M}}

\newcommand{\sI}{{\mfs I}}
\newcommand{\sO}{{\mfs O}}

 \usepackage{braket}
\newcommand{\N}{\mathbb{N}}
\newcommand{\C}{\mathbb{C}}
\newcommand{\R}{\mathbb{R}}

\begin{document}

\title{Unitarity and information in quantum gravity: a simple example}

\author{Lautaro Amadei}
\affiliation{{Aix Marseille Univ, Universit\'e de Toulon, CNRS, CPT, Marseille, France}}

\author{Hongguang Liu}
\affiliation{{Aix Marseille Univ, Universit\'e de Toulon, CNRS, CPT, Marseille, France}}

\author{Alejandro Perez}
\affiliation{{Aix Marseille Univ, Universit\'e de Toulon, CNRS, CPT, Marseille, France}}

\date{\today}

\begin{abstract}
In approaches to quantum gravity, where smooth spacetime is an emergent approximation of a discrete 
Planckian fundamental structure, any effective smooth field theoretical description would
miss part of the fundamental degrees of freedom and thus break unitarity. This is applicable also to trivial  gravitational field (low energy) idealizations realized by the use of the Minkowski background geometry which, as any other spacetime geometry, corresponds, in the fundamental description, to infinitely many different and closely degenerate discrete microstates. The existence of such microstates provides a large reservoir for information to be coded at the end of black hole evaporation and thus opens the way to a natural resolution of the black hole evaporation information puzzle. 

In this paper we show that these expectations can be made precise in a simple quantum gravity model for cosmology motivated by loop quantum gravity. Concretely, even when the model is fundamentally unitary, when microscopic degrees of freedom irrelevant to low-energy cosmological observers are suitably ignored, pure states in the effective description evolve into mixed states due to decoherence with the Planckian microscopic structure. Moreover, in the relevant physical regime these hidden degrees freedom do not carry any `energy' and thus realize in a fully quantum gravitational context the idea (emphasized before by Unruh and Wald) that decoherence can take place without dissipation, now in a concrete gravitational model strongly motivated by quantum gravity.  All this strengthen the perspective of a quite conservative and natural resolution of the black hole evaporation puzzle where information is not destroyed but simply degraded (made unavailable to low energy observers) into correlations with the microscopic structure of the quantum geometry at the Planck scale.

\end{abstract}
\pacs{98.80.Es, 04.50.Kd, 03.65.Ta}

\maketitle
\tableofcontents

%

\definecolor{mycolor}{rgb}{0.122, 0.435, 0.698}

\section{Introduction}

The mathematical models that so far define our successful physical theories are all reversible in the sense that they can predict the future value of the variables they use from their initial values, while conversely the past can be uniquely reconstructed from the values of these variables in the future. The memory of the initial condition is not lost in the dynamics and their information content remains. This is true for classical mechanics and field theory and it is also true for quantum mechanics and quantum field theory as long as we do not invoke the postulate of the collapse of the wave function (i.e. as long as we do not intervene from the outside via a measurement \footnote{This is not the case in modifications  of quantum mechanics where the collapse of the wave function happens spontaneously. In such theories information is actually destroyed (for a discussion of black hole evaporation in such contexts see \cite{Modak:2014vya, Okon:2016qlh, Okon:2017pvc}).}). In the quantum mechanical  setting this property boils down to the fact that evolution to the future is given by a unitary operator which can always be undone via its adjoint transformation.

This property of our fundamental models has always troubled naive intuition when faced with situations that appear to be irreversible. For example: what would happen with these words if the computer collapses at this very moment. What if after printed this paper is burned. Common sense would answer that the information in these pages (if of any relevance)  would be lost. However, the physicist, trained to firmly believe in the statement of the previous paragraph, would say that the information in these words is not lost but simply hidden (to the point of becoming unrecoverable) in the humongous number of microscopic variables that would describe the whole system. In the case of burning the paper, these words remain `written' (it would be claimed) in the multiple correlations between the degrees of freedom of the molecules in the gas of the combustion diffusing in the atmosphere while transferring the information to even larger and yet pristine portions of the very large phase space of an unbounded universe. In the case where the computer collapses, a similar story can be told involving the dissipation of the bits into the environment. Of course the physicist cannot prove this; however, it is a consistent story in view of the strongly cherished principle of unitarity.   

Such effective irreversibility is clearly captured in the second law of thermodynamics stating that (for suitable situations involving large number of degrees of freedom) entropy can only increase. At the classical level this clashes at first sight with the Liouville theorem stating that the phase space volume of the support of  a distribution in phase space is preserved by dynamical evolution. However, nothing restricts the shape of this volume to evolve into highly intricate forms that a macroscopic observer might be unable to resolve. More precisely,  suitable initial conditions that the observer agent regards as special (for instance the macroscopic configurations of ink particles defining words in this paper before the fire reached them) come with an  uncertainty in accordance to the observers limited measurement capabilities. This is idealized by a distribution in phase space occupying an initial phase space volume of a regular shape (these ensemble of points represent the system in what follows). Now as time goes by the apparent phase space volume (not the real volume which remains constant) would seem to grow to the agent just because of its intrinsic inability to separate the points in phase space that the systems occupies from the close neighbouring ones where the system is not.  In this sense the arrow of time is only emergent macroscopically due to the special initial conditions, and the intrinsic coarse graining introduced by a macroscopic observer with its limitations. We will argue that the general lines of this story remain the same when black hole evaporation is considered.

General relativity combined with quantum field theory, in a regime where both are expected to be good approximations, imply that large isolated black holes behave like thermodynamical systems in equilibrium. They are objects close to equilibrium at the Hawking temperature that lose energy extremely slowly via Hawking radiation.  When perturbed they come back to equilibrium to a new state and the process satisfies the first law of thermodynamics with an entropy equal to $1/4$ of the area $A$ of the black hole horizon in Planck units. Under such perturbation (which in particular can be associated also to their slow evaporation) the total entropy of the universe can only increase, namely 
\be  
\delta S=\delta S_{\rm matter}+\frac{\delta A}{4}\ge 0, 
\ee
where $\delta S_{\rm matter}$ represents the entropy of whatever is outside the black hole (including for instance the emitted radiation).

This quasi equilibrium phase (which can be extremely long lasting for macroscopic black holes) is only an intermediate situation before complete evaporation. This intermediate phase is predicted by general relativity as the result of gravitational collapse which takes place for suitable initial conditions. The point is that the irreversibility captured by the previous equation can once more be associated to the same ingredients present in our previous example: the special nature of the initial conditions in view of our biased criterion of macroscopic observers (low curvature and low densities in the past), high curvature and a huge new phase available in the Planckian regime near what would be the singularity in general relativity (the {\em would-be-singularity} from now on). As emphasized by Penrose (see for instance \cite{penrose-roadtoreality-2005}),  among others, the arrow of time comes from the fact that we started with a spacetime that was well approximated by a low curvature one with some dilute matter distribution (gas and dust) that would first form  large stars that one day can collapse to form black holes\footnote{To these two speciality conditions one might also have to add one concerning the state of the hypothetical microscopic constituents at the Planck scale if the view we are advocating here and in \cite{Perez:2014xca, Perez:2017cmj} is correct.}.   Before  the formation of a black hole  the story of our system exploring larger and larger portions of the available phase space is the usual and standard one involving molecules, atoms and fundamental particles. The perspective we want to stress here is that the story continues to be the same after the black hole forms, but now a new and huge new portion of phase space has is opened by the gravitational collapse: the internal {\em would-be-singularity} of the classical description beyond the event horizon. Like the lighter setting the paper in fire and allowing for fast chemical reactions that degrade the ink in these words when burning the paper, the singularity brings the system in contact with the quantum gravity scale. The gravitational collapse ignites interactions with the Planckian regime inside the black hole horizon (see Footnote \ref{fufu}), and that must be (as in the burning paper) the key point for resolving the puzzle of information in black hole evaporation. This perspective was advocated in \cite{Perez:2014xca, Perez:2017cmj}.

\begin{figure}[h] \centerline{\hspace{0.5cm} \(
\begin{array}{c}
\includegraphics[width=9cm]{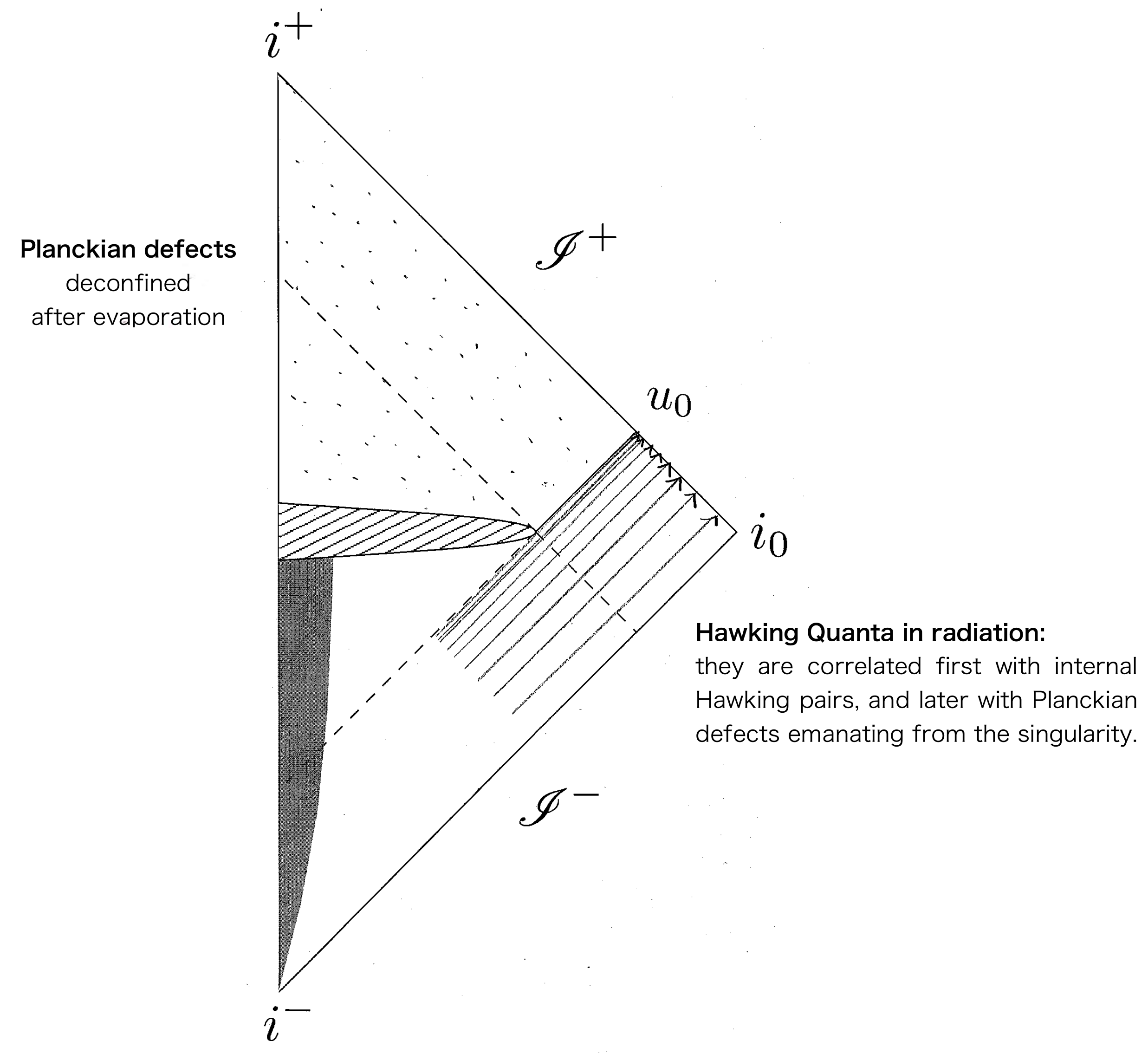} \end{array}\ \ \begin{array}{c}
 \includegraphics[width=9cm]{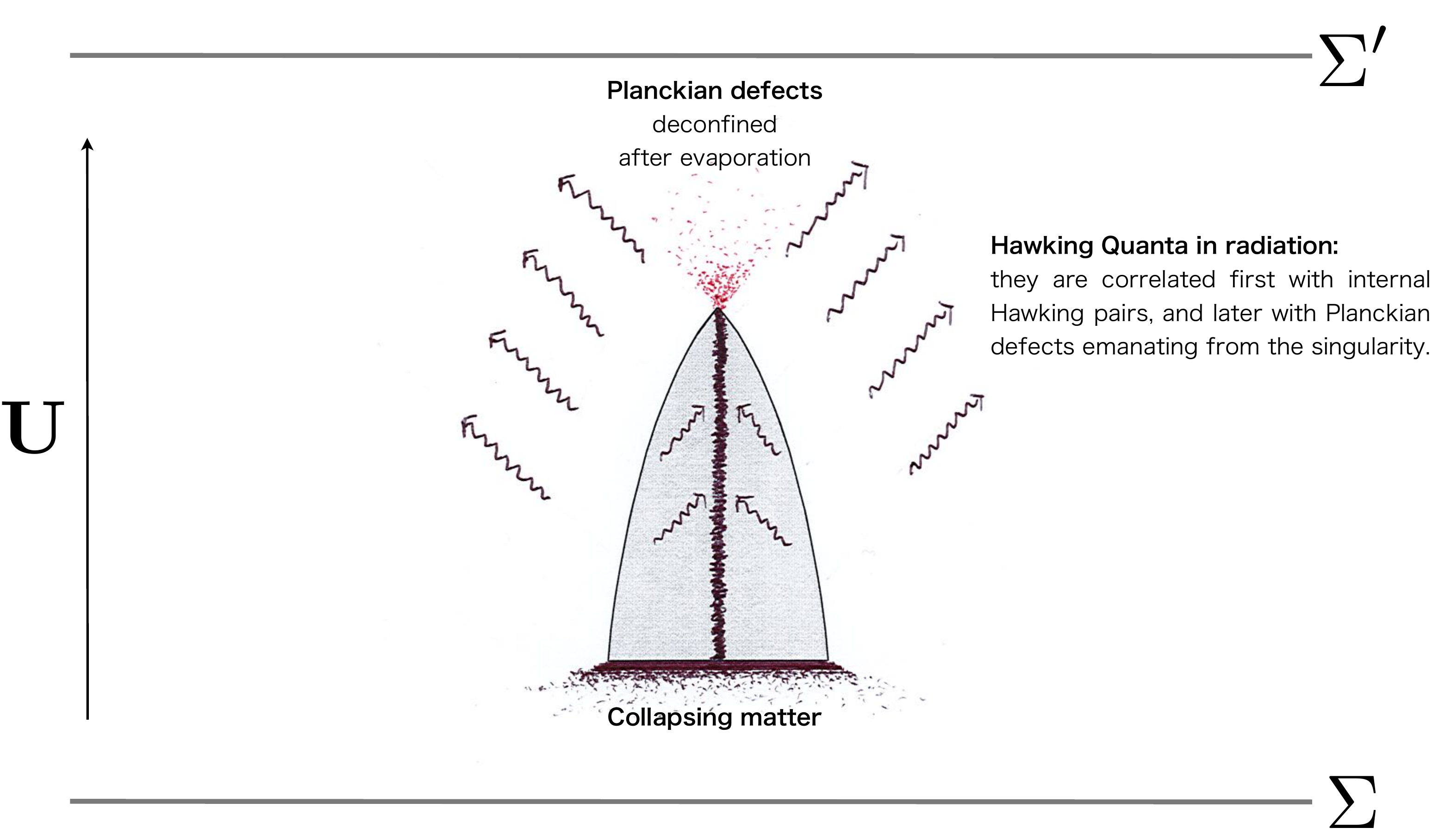}
\end{array}
\)}
\caption{{\bf Left panel:} Penrose diagram illustrating (effectively) the natural scenario, suggested by the fundamental features of LQG, for the resolution of the information puzzle in black hole evaporation \cite{Perez:2014xca}. The shaded region represents the {\em would-be-singularity} where high fluctuations in geometry and fields are present and where the low energy degrees of freedom of the Hawking pairs are forced to interact with the fundamental Planckian degrees of freedom. {\bf Right panel:} Same situation as a scattering process from an initial to a final Cauchy hypesurface. This figure contains basically the same information as the Penrose diagram. Its additional merit, if any, is the more intuitive representation of the shrinking black hole as well as the time scales involved (collapse time is very short with respect to the evaporation time). Both these features are absent in the conformal representation on the left. There are limitations of such spacetime representation of a process that is fundamentally quantum and hence only understandable in terms of superpositions of different spacetime geometries.}
\label{funo}
\end{figure}

Let us briefly describe the scenario introduced in these papers  with the help of Figure \ref{funo}. 
The first assumption in the diagram is that there is evolution across the {\em would-be-singularity} (predicted by the classical dynamics) inside the black hole. This assumption is intrinsic in the representation in the figure; however,  the scenario still makes sense if instead a baby universe is formed, i.e., if the {\em would-be-singularity} remains causally disconnected from the outside after evaporation.  In that case the correlations established with the baby universe remain hidden forever to outside observers. The virtue of the present scenario in such case  would be to give an identity to the degrees of freedom involved. The idea that the spacetime representation of the situation resembles the one in Figure \ref{funo} comes from the various results in symmetry reduced models for both cosmology \cite{Bojowald:2001xe, Ashtekar:2006wn} as well as for black holes \cite{Modesto:2004xx, Modesto:2005zm, Bojowald:2005qw, Bojowald:2005ah, Gambini:2013ooa, Corichi:2015xia, Ashtekar:2018lag} and was first pictured in \cite{Ashtekar:2005cj}. In such a context a `scattering theory' representation (where an in-state evolves into an out-state) is possible even though the result (as we will argue) cannot be translated into the language of effective quantum field theory.

But what do we mean by a black hole in this evaporating context? In the asymptotically flat idealization, the black hole region is defined in classical general relativity as the portion of the spacetime $\sM$ that is not part of the past of $\sI^+$. Such definition needs to be modified in quantum gravity.  In order to do that we introduce the notion of the semiclassical past $J_{\va \rm C}^{-}(\sI^+)$ of $\sI^+$ as the collection of events in the spacetime that can be connected to $\sI^+$ by causal curves along which the Kretschmann scalar $K\equiv R_{abcd}R^{abcd}\le {\rm C} \ell_p^{-4}$ for some constant $\rm C$ of order unity. The black hole region can be defined then as\be B \equiv \sM- J_{\va \rm C}^{-}(\sI^+). \ee
Its dependence on the constant $\rm C$ is not an important limitation in the discussion about information. Different $\rm C$ would lead to BH regions that coincide up to Planckian corrections.

The most clear physical picture emerges from the analysis of the Penrose diagram on the left panel of Figure \ref{funo}  from the point of view of observers at future null infinity $\sI^+$. These observers are assumed to be at the center of mass Bondi frame of the BH formed via gravitational collapse. We also assume that the Bondi mass of the BH is initially $M\gg m_p$ at some retarded time $u$ on $\sI^+$ representing the time where the BH has achieved its quasi-equilibrium state and starts evaporating slowly via Hawking radiation, i.e., the BH is initially macroscopic. Under such conditions the evaporation is very slow and we can trust the semiclassical description that tell us that the Bondi mass $M(u)$ will slowly decrease with $u$ from this initial value $M$ until times $u=u_0$ (see figure) with $M(u_0)\gtrapprox m_p$ in a time of the order of $M^3$ in Planck units.  From this time on the details depend on a full quantum gravity calculation because the curvature around the BH horizon has become Planckian. Nevertheless, independently of such details we can safely say that the spacetime and the matter degrees of freedom encoded on $\sI^+$ for $u>u_0$ must be in a superposition of states all of which are very close to flat space-time, as far as their geometry is concerned, with matter excitations very close to the vacuum because there is only at best an energy of the order of $E_{\rm \va late}\approx m_p$  to substantiate both. In addition these excitations  must be correlated with the early Hawking radiation with energy $E_{\rm \va early}\approx M-m_p$ if unitarity is to hold. The late degrees of freedom are often referred to as {\em purifying degrees of freedom}.

One possibility is to assume that such purifying degrees of freedom are particle excitations coming from what is left of the BH (a remnant). Now, because these particles must be extremely infrared because only 
 $E_{\rm \va late}\approx m_p$ is available for purification, then a simple estimate of the time (denoted $\tau_p$) that the process would have to last if this is the main channel for purification yields $\tau_p\approx (M/m_p)^4$. This is the scenario of an extremely long lasting point-particle-like remnant with a huge internal degeneracy which is claimed to be problematic from the point of view of effective quantum field theory \cite{Banks:1992is}.   

The possibility we put forward is that if smooth spacetime is an emergent notion from an underlying discrete physics, then the classical geometries of general relativity with quantum fields living on them would only restrict the fundamental Hilbert space to a subset containing very large (possibly infinite) number  of states. For instance the Minkowski vacuum unicity in standard quantum field theory would fail in the sense that the requirement that states look {\em flat}  for (coarse grained) low energy observers---which are those for which an effective quantum field theory description in terms of smooth fields living on a smooth geometry is a suitable approximation---would still admit highly denegerate ensemble (all with total mass indistinguishable from zero by these observers). Now, such low energy modes cannot be interpreted as the infrared excitations of fields mentioned in the previous paragraph (say low energy photons) because they are smooth low energy configurations. These low energy degrees of freedom would correspond to defects in the Planckian fabric of quantum gravity which we simply are not sensitive to with our coarse low energy probes (like the molecular structure that escapes the smooth characterization of the Navier-Stokes effective theory of fluids).

Why should information be hidden in the UV and not be IR modes as in the remnant scenario mentioned above? It is often believed that because the volume inside the black hole actually becomes very large (according to suitable definitions \cite{Christodoulou:2016tuu}) then modes that are correlated with the Hawking radiation are redshifted and become highly IR inside. Although this is true for spherically symmetric Hawking quanta in the spherically symmetric Schwarzshild background---where such modes are indeed infinitely redshifted as detected by regular observers when they approach the singularity at $r=0$---this conclusion fails when one considers no-spherical modes no matter how small the deviation from spherical symmetry is \footnote{\label{fufu} In the Schwarzshild background, the frequency measured by a radially freely falling observer normal to the $r=$constant hypersurfaces goes like
\be
\omega^2(r)=\frac{\ell^2}{r^2}+\frac{r}{2M} {\cal E}^2 +\sO\left(\frac{r^2}{M^2}\right),
\ee
%
%
where ${\cal E}=-k\cdot \xi$ and $\ell=k\cdot \psi$ are the conserved quantities associated to the massless particle with wave vector $k^a$ and $\xi^a$ and $\psi^a$ are the stationarity and rotation killing fields of the background. The qualitative behaviour approaching $r=0$ would be the same for any other observer measuring $\omega$ (the divergence of $\omega$ is observer independent). Only exactly spherically symmetric modes with $\ell=0$ would become IR at the singularity. However, this conclusion is no longer true if the BH rotates or if we consider that at the fundamental level states with exact spherical symmetry inside the BH are of measure zero. Notice that such UV divergence in the non-spherically symmetric Hawking partners implies large deviations from spherical symmetry near the singularity (if their back reaction would be taken into account). This should be kept in mind when modelling the situation with spherically symmetric mini-superspace quantum gravity models.}. Therefore, generically all modes become UV close to the singularity.     

We can draw a formal analogy with the Unruh effect  as follows. The Unruh effect arises from the structure of the vacuum state $\ket{0}$ of a quantum field on Minkowski spacetime when written in terms of the modes corresponding to Rindler accelerated observers with their intrinsic positive frequency notion. The vacuum takes the form
\be
\ket{0}=\prod_{k}\left( \sum_n \exp\left({-n\frac{\pi \omega_k}{a}}\right) \ket{n,k}_{R}\otimes\ket{n,k}_{L}\right),
\ee
where $\ket{n,k}_{L}$ and $\ket{n,k}_{R}$ define the particle modes---as viewed by an accelerated observer with uniform acceleration $a$---on the left and the right of the Rindler wedge
\cite{Wald:1995yp}.
Here we see from the form of the previous expansion that even when we are dealing with a pure state (if we define the density matrix $\ket{0}\bra{0}$), the reduced density obtained by tracing over one of the two wedges would produce a thermal state with $T=a/(2\pi)$. The statement in the perspective we propose on the purification of information in   
BH evaporation  can we schematically represented (the following is certainly not a precise equation) by  
\be
\mathbf{U}\!\!\!\!\!\!\!\!\!\underbrace{\ket{\rm flat,0}}_{\rm quantum \ geometry}\!\!\!\!\!\!\!\!\! \otimes\!\!\!\!\!\!\!\!\! \overbrace{\ket{\phi}}^{\rm matter\ fields}\!\!\!\!\!\!\!\!\!=\prod_{k}\left( \sum_n \exp\left({-\frac{\beta}{2} n \omega_k}\right) \ket{\rm flat,n} \otimes\ket{n,k}\right),
\ee
where an initial state of flat quantum geometry $\ket{\rm flat, 0}$ tensor a state representing  initially diluted matter fields $\ket{\phi}$ evolves unitarily via $\mathbf U$ into the formation of a BH and the subsequent evaporation (Figure \ref{funo}) which after complete evaporation is written as a superposition of flat quantum geometry states $\ket{\rm flat, n}$---which are all indistinguishable from $\ket{\rm flat, 0}$ to low energy agents and differ among them by quantum numbers $n$ corresponding to quantities that are only measurable if one probes the state down to its Planckian structure---tensor product with normal $n$-particle excitations of matter fields representing Hawking radiation. As mentioned above the previous equation is only schematic. Is main inappropriateness is the fact that the reduce density matrix obtained by tracing over the quantum geometry hidden degrees of freedom would give a thermal state at a fixed temperature $T$. This is at odds with the expectation that the Hawking radiation should contain a superposition of the thermal radiation  emitted at different Hawking temperatures during the long history of the evaporation of the BH.   But the point that this equation and the discussion of the previous paragraph should make clear is that the purification mechanism proposed here has nothing to do with the point like remnant scenario with all its problems associated to a long lasting particle-like remnant.  Here, to the future of the {\em would-be-singularity} in Figure \ref{funo}, we simply have a quantum superposition of different quantum geometry states that all look flat to low energy observers. There are no localized remnant hiding the huge degeneracy inside; there is only a large superposition  of states that are inequivalent in the fundamental quantum gravity Hilbert space but seem all the same for low energy agents. Such degrees of freedom cannot be captured by any effective description in terms of smooth fields (EQFT) for the simple reason that they are discrete in their fundamental nature.

 \begin{figure}[h] \centerline{\hspace{0.5cm} \(
\begin{array}{c}
\includegraphics[width=14cm]{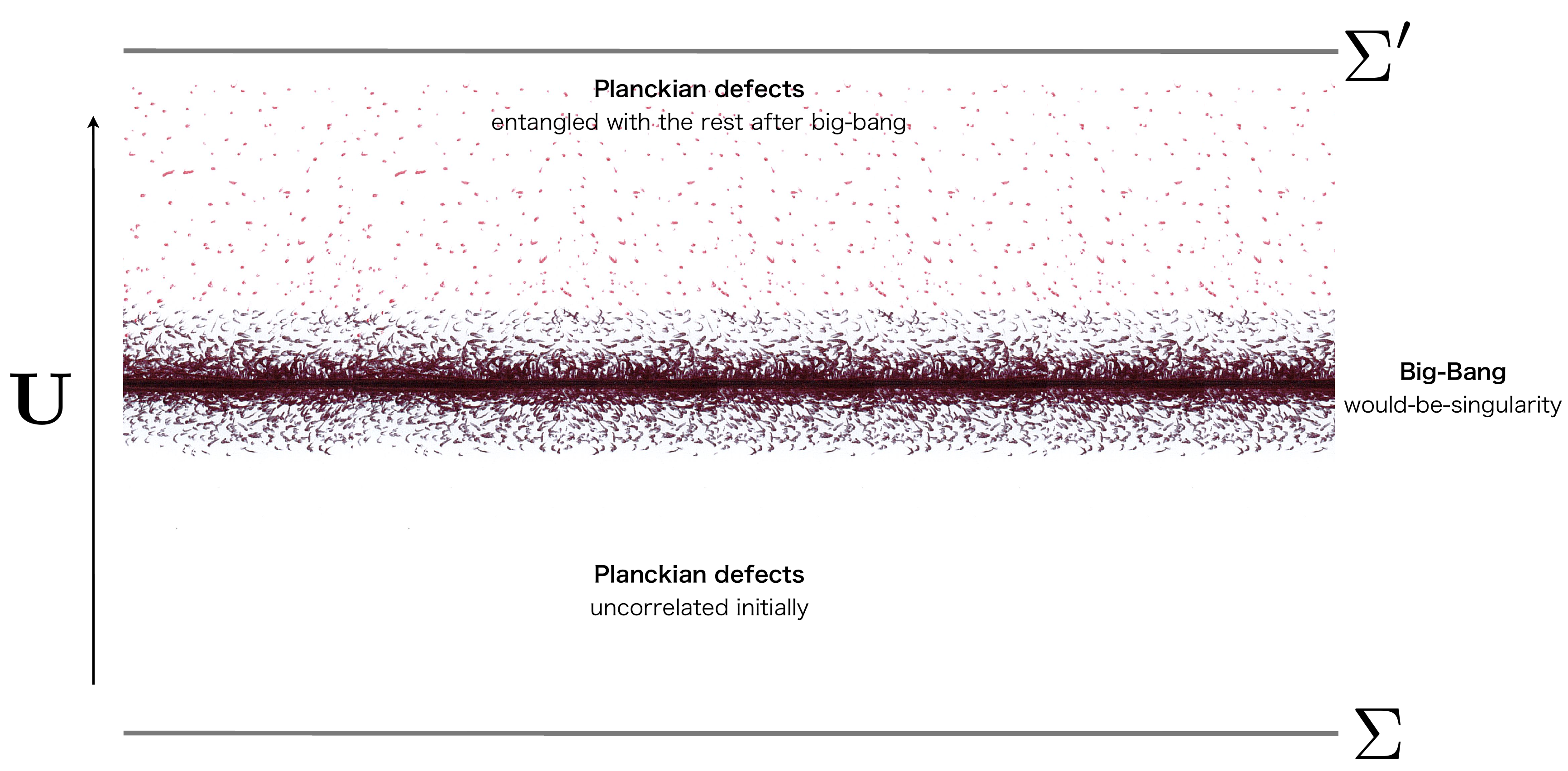}
\end{array}
\)}
\caption{Diagram illustrating (effectively) the natural scenario, suggested by the fundamental features of LQG, for the resolution of the information puzzle in black hole evaporation \cite{Perez:2014xca}. As in Figure \ref{funo}, one should keep in mind the limitations of such spacetime representation of a process that is fundamentally quantum and hence only understandable in terms of superpositions of different spacetime geometries.}
\label{fdos}
\end{figure}

Notice that the degrees of freedom where information would be coded after BH evaporation do not satisfy the usual Einstein-Planck relationship $E=\hbar \omega$ or equivalently $p=h/\lambda$ (for some `wavelength' $\lambda$ or `frequency' $\omega$) and this might deceive intuition \footnote{A nice counter example of this intuition is given by the case of a non relativistic charged particle in a two dimensional infinite perfect conductor in a uniform magnetic field normal to the conducting plane. The energy eigenvalues are given by the Landau levels $E_n=\hbar \omega_{\rm B} (n+1/2)$ where $\omega_{\rm B}=qB/(mc)$ is the Bohr magneton frequency, but they are infinitely degenerate. There are canonically conjugated variables $(P,Q)$ associated to the particle that are cyclic, i.e., do not appear in the Hamiltonian. In this case one can produce wave packets that are as `locallized' as wanted in the variable $Q$ without changing the energy of the system. Interestingly, this is a perfect example of system where one could have an apparent loss of information of the type we are proposing here (for a more realistic analog gravity model discussing the information paradox along the lines of the present scenario see \cite{Liberati:2019fse}). If one scatters a second particle interacting softly with the charged particle on the plate so that the interaction does not produce a jump between different Landau levels, then correlations with the cyclic variables would be established without changing the energy of the system. This is the perfect model to illustrate the possibility of decoherence without dissipation. } . They are described as Planckian defects nevertheless they do not carry Plankian energy. The point is that such relationship only applies under suitable conditions which happen to be meet in many cases but need not be always valid.
One case is the one of degrees of freedom that can be thought of as waves moving on a preexistent spacetime. This is the case of particle excitations in the Fock space of quantum field theory or effective quantum field theories which are defined in terms of a preexistent spacetime geometry. There is no clear meaning to the above intuitions in the full quantum gravity realm where the present discussion is framed.  Even when such relations (linked to the usual uncertainty principle of quantum mechanics) should hold in a suitable sense---if the structure suggested by canonical quantization survives in quantum gravity as it should to a certain degree---they would apply to phase space variables encoding the true degrees of freedom of gravity that we expect (from the general covariance of general relativity) to be completely independent of a preexistent background geometry.  We will see that such degrees of freedom with such peculiar nature actually arise naturally in the toy model of quantum gravity that we analyze in this article.

It is presently hard to prove that such scenario is viable in a quantum theory of gravity simply because there is no such theoretical framework developed enough for tackling BH formation and evaporation in detail. However, the application of loop quantum gravity to quantum cosmology leads to a model with similar features, where evolution across the classical singularity is well defined \cite{Bojowald:2001xe}.  The results have been reported in \cite{Amadei:2019ssp}. In this article we present the main features of this model in more detail and show that the conclusions of \cite{Amadei:2019ssp}, some of which drawn from some simplified modelization of the matter matter coupling, are generic and remain true in more physically realistic models. The results can be described briefly by making reference to the Figure \ref{fdos} which should be compared with the right panel in Figure \ref{funo}. We will show that the evolution in loop quantum cosmology from a universe in an initially contracting state in the past of the {\em would-be-singularity} to an expanding universe in its future is perfectly unitary in the fundamental description. Nevertheless, states in the Hilbert space of loop quantum cosmology contain quantum degrees of freedom which are hidden to low energy coarse grained observers. If these degrees of freedom are traced out of the initial density matrix then we will see that pure states (in the sense of the reduced density matrix) generically evolve into mixed states across the {\em would-be-singularity}. Information is lost into correlations with degrees of freedom that are Planckian and thus inaccessible to macroscopic observers. These correlations are established in an inevitable way during the strong curvature phase of evolution across the big bang (just as expected in the BH scenario described above). As energy is conserved  (energy is a delicate notion in cosmology but happens to be well defined in our model as we will see) the defects that purify the final state do not enter in the energy balance which realizes another crucial necessary ingredient of the general scenario (decoherence happens with negligible dissipation \cite{Unruh:2012vd}). 

Let us finish this introduction with a very brief description of the organization of the paper. The paper is basically separated in two parts. In the first part (Section \ref{part1}) we show that the scenario we have described is realized in unimodular quantum cosmology following the standard quantization prescription of loop quantum cosmology. The model of this section corresponds exactly with the type of models in the standard literature \cite{Ashtekar:2011ni}. In the second part of the paper we observe that there is natural extension of loop quantum cosmology based on the regularization ambiguity associated with the quantization of the Hamiltonian. This extension opens a new chanel for information to flow. Although this second option is not necessary to illustrate our point (already realized in the standard theory in the first part) it gives a different identity to the defects which could lead to independent and thus useful insights. There are a series of apendices where some calculations are done. Appendix \ref{scalarf} is specially important because some of the over simplified model (analytic) calculations in the body of the paper are done (numerically) in the more physically realistic case of the coupling of gravity with a massless scalar field.

\section{Unimodular Gravity: four volume as time}\label{part1}

Unimodular Gravity is nearly as old as General Relativity itself, it was introduced by Einstein in 1919 \cite{Einstein1919Spielen-Gravita} as an attempt to describe nuclear structure geometrically. In this work Einstein identifies also an appealing feature of the theory which is the fact that the cosmological constant arises as a dynamical constant of motion that needs to be added to the initial values of the theory. In unimodular gravity the cosmological constant is a constant of integration and not a universal or fundamental constant of nature. Interest in the theory was regained in the late 80's with the observation of Weinberg \cite{Weinberg:1988cp} that, for the above reason, semiclassical unimodular gravity provides a trivial resolution of the cosmological constant problem as vacuum energy simply does not gravitate.  Unimodular gravity is the natural low energy description that emerges from the formal thermodynamical ideas of Jacobson \cite{Jacobson:1995ab} and represents the expected low energy regime of the causal set approach \cite{Bombelli:1987aa}.  

Another property of unimodular gravity (specially important for us here) is that it completely resolves the problem of time \cite{Unruh:1988in} in the cosmological FLRW context. More precisely, the theory comes with a preferred time evolution and a preferred Hamiltonian (the energy of the universe is well defined and directly linked with the value of the cosmological constant). The quantum theory is described by a Schroedinger like equation where states of the universe are evolved by a unitary evolution operator. Therefore, unlike the general situation in quantum gravity, the notion of unitarity is unambiguously defined in unimodular quantum cosmology. This is the main reason why unimodular gravity provides the perfect framework for the discussion of the central point of this work.

Here we specialize to homogeneous and isotropic cosmologies that are spatially flat (k=0), i.e.,  the spatial manifold $\Sigma$ is topologically $\mathbb{R}^{3}$. What follows is the standard construction. For a detailed account of the Hamiltonian analysis in the cosmological framework see \cite{Chiou:2010ne} . The FLRW metric is
\be
ds^2=-N(t)^2 dt^2+a(t)^2 \underbrace{(dx^2+dy^2+dz^2)}_{\mathring q_{ab}},
\ee
where $\mathring q_{ab}$ denotes the fiducial spacial metric. Since $\Sigma$ is non-compact some  integrals are infrared divergent and are regularized by restricting to a fixed fiducial cell $\mathcal{V}$ of fiducial volume $V_{0}$ with respect to the fiducial spacial metric 
\begin{equation}\label{eq10}
    \mathring q_{ab} =  \mathring e^{i}_{a}  {\mathring e^{j}_{b}} \ \delta_{ij},
\end{equation}
where $ {\mathring e^{i}_{a}} $ denotes a fiducial triad and the physical metric is given by $q_{ab} = a^{2}(t ) { \mathring q_{ab}}$.
The action of unimodular gravity in the FLRW minisuperspace setup  is given by
\ba\label{adm_action_cosmo}
 S[a,\dot{a}, \lambda] &=& \frac{3}{8 \pi G}\int_{\mathbb{R}}  \left(  \frac{V_{0} a \dot{a}^{2}}{N}+\lambda V_0 (N a^3-1) \right)dt,
\ea    
where $\lambda$ is a Lagrange multiplier imposing the unimodular constraint $N=a^{-3}$ (i.e. $\sqrt{|g|}=1$), and  the first term is the Einstein-Hilbert action restricted to the FLRW geometries \footnote{There is an overall minus sign the definition action with respect to standard treatments. This is done so that the the pure-geometry Hamiltonian is positive definite.}. In order to use loop quantum cosmology techniques (for a discussion of the quantization in the full loop quantum gravity context see \cite{Smolin:2010iq, Smolin:2009ti}) one introduces the new canonical variables $c$ and $p$ via the basic Ashtekar-Barbero connection variables $A^{i}_{a}$ and $E^{a}_{i}$, namely   
\ba\label{eq11}
    E_{i}^{a} =
p \, \left({\mathring e_{i}^{a}} V_{0}^{-2/3} \right),  \ \ \ \ \ \ \ \ \ \ \ \ \ 
    A_{a}^{i} =
    c \,   \left( {\mathring \omega_{a}^{i}}V_{0}^{-1/3}\right),
\ea
where ${\mathring \omega_{a}^{i}}$ is a fiducial reference connection.
These variables are related to those in \eqref{adm_action_cosmo} via the equations
\begin{equation}\label{eq33}
    |p|= V_{0}^{2/3} a^{2},  \qquad     c  = V_{0}^{1/3} \frac{\gamma \dot{a}}{N}.
\end{equation}
The action becomes
\be
S[c,p,\lambda]=\frac{3}{8\pi G }\int_{\R}
\gamma^{-1}c \dot p-N \gamma^{-2} \sqrt{|p|} c^{2}
+\lambda (N |p|^{\frac 32}-V_0) dt,
\ee
 and  $c$ and $p$ are canonically conjugated  in the sense that \begin{equation}\label{eq12}
    \{ c,p \}  = \frac{8 \pi G \gamma}{3}.
\end{equation}
The unimodular condition $N=a^{-3}$ fixes the lapse to $N=V_0/|p|^{3/2}$  and the unimodular Hamiltonian becomes 
\be\label{mavi}
H=\frac{3 V_0}{8\pi G} \frac{c^2}{\gamma^2 |p|}.
\ee
The proportionality of the Hamiltonian with $V_0$, and the fact that the four volume bounded by $V_0$ at two different times is given by $v^{\va(4)}=V_0 \Delta t$, implies that  time evolution can be parametrized in terms  the four volume elapsed from some reference initial slice. The associated Hamiltonian (conjugated to $v^{\va(4)}/(8\pi G)$) is 
\be\label{cosmoco}
\Lambda=\frac{3 c^2}{\gamma^2 |p|},
\ee
and corresponds to the cosmological constant.

\subsection{Quantization}\label{qqq}
The loop quantum cosmology quantization uses a non standard representation of the canonical variables where the variable $c$ does not exist as a quantum operator, and the definition of the Hamiltonian requires a special regularization procedure known as the $\bar\mu$-scheme \cite{Ashtekar:2011ni}. The quantization prescription is greatly simplified by the introduction of new canonically conjugated dynamical variables $b$ and $\nu$ defined as \cite{Ashtekar:2008jd}
\begin{equation}\label{eq38}
    b \equiv \frac{c}{|p|^{\frac12}} \qquad \nu \equiv {\rm sign}(p) \frac{|p|^{\frac32}}{2 \gamma \pi \ell_p^2},
\end{equation}
with Poisson brackets \footnote{\label{footy}The factor $\hbar^{-1}$ appears on the right hand side of the Poisson brackets due to the introduction of $\hbar$ (via $\ell_p^2$) in the definition of the new variable $\nu$. This is done to match standard definitions \cite{Ashtekar:2011ni}.}
\begin{equation}\label{eq39}
    \{ b,\nu \} = 2\hbar^{-1}.
\end{equation}
The variable $\nu$ corresponds to the physical volume of the fiducial cell divided by $\ell_p^2$; it has units of distance. The variable $b$ is simply its conjugate momentum. 
In terms of these variables the gravitational (unimodular) Hamiltonian (\ref{mavi}) integrated in a fiducial cell $\cal V$ becomes
\begin{equation}\label{eq40}
    H= \frac{3 V_0}{8\pi G \gamma^2}  b^{2} 
\end{equation}
Note the extreme simplicity of the previous expression: the unimodular hamiltonian is just the analog of that of a free particle in one dimension with mass parameter $m=4 \pi \gamma^2/(3 V_0) $ and momentum $b$. In the absence of matter, the Hamiltonian can be quantized in the Wheeler-DeWitt representation where the evolution in unimodular time is unitary and there is no singularity (the classical solutions correspond to De-Sitter geometries with arbitrary but positive cosmological constants). 
The singularity in the classical theory becomes real when matter is introduced. 

In the loop quantum cosmology polymer representation, just as for $c$, there is no operator corresponding to $b$  but only the operators corresponding to finite $\nu$ translations \cite{Ashtekar:2006wn}; from here on referred to as shift operators \be\label{shifty} {\exp(i 2 k b)} \triangleright\Psi(\nu )=\Psi(\nu-4k).\ee 
For $k=q\sqrt{\Delta} \ell_p$ and $q\in \N$, states that diagonalize the previous shift operators, denoted $\ket{b_0; \Gamma^{\epsilon}_\Delta}$, are labelled by a real value $b_0$ and by a graph $\Gamma^{\epsilon}_\Delta$. The graph is a 1d lattice of points in the real line of the form $\nu=4n \sqrt{\Delta}\ell_p +\epsilon$ with $\epsilon\in [0,4 \sqrt{\Delta}\ell_p)$ and $n\in \N$. The corresponding  wave function is given by $\Psi_{b_0}(\nu)\equiv \braket{\nu|b_0; \Gamma^{\epsilon}_\Delta}=\exp{(-i\frac{b_0 \nu}{2})} \delta_{\Gamma^{\epsilon}_\Delta}$ where the symbol $\delta_{\Gamma^{\epsilon}_\Delta}$ means that the wavefunction vanishes when  $\nu\notin \Gamma_\Delta^\epsilon$. It follows from (\ref{shifty}) that
\be\label{eige}  {\exp(i 2 k b)} \triangleright \ket{b_0; \Gamma^{\epsilon}_\Delta}=
\exp{(i2 k b_0)} \ket{b_0; \Gamma^{\epsilon}_\Delta}.\ee
 The states $\ket{b; \Gamma^{\epsilon}_\Delta}$ are eigenstates of the shift operators that preserve the lattice $\Gamma^{\epsilon}_\Delta$. Notice, that the eigenvalues are independent of the parameter $\epsilon$. i.e. they are infinitely degenerate and span a non separable subspace of the quantum cosmology Hilbert space $\sH_{lqg}$.

A scale $\bar\mu$ is needed in order to define a regularization of \eqref{eq40} representing the Hamiltonian in $\sH_{lqc}$. The reason is that there is no operators associated to $b$ but only approximants constructed via the shift operators (\ref{shifty}). The so-called $\bar \mu$-scheme \cite{Ashtekar:2011ni} introduces a dynamical length scale $\bar \mu$ defined as
\begin{equation}\label{eq17}
    \overline{\mu}^{2} = \frac{\ell_{p}^{2}\Delta }{|p|},
\end{equation}
where $\Delta$ represents the so-called `area-gap' which plays the role of a UV regulator. It is normally associated to the smallest non-vanishing area quantum in the full theory of loop quantum gravity.  For the moment (as in the standard treatement) this is just a fixed parameter\footnote{In Section \ref{maco}, we will turn this quantity into a quantum operator acting on the microscopic sector of the Hilbert space that will be introduced.}. When translated into the variables \eqref{eq38} $\overline\mu$ corresponds to considering approximants to $b$ constructed out of shift operators \eqref{shifty} with fixed $k\equiv \sqrt{\Delta} \ell_p$. In terms of these one obtains the following regularization of the Hamiltonian (\ref{eq40}) which is a well defined self-adjoint operator\footnote{The Hamiltonian $\hat{H}_{0}$ \eqref{eq89b} is symmetric,  that is $\braket{\Psi_{1},\hat{H}_{0} \Psi_{2}} = \braket{\hat{H}_{0}\Psi_{1}, \Psi_{2}}$, with respect to the inner product $
    \braket{\Psi_{1},\Psi_{2}} = \sum_{\nu} \overline{\Psi_{1}(\nu)} \Psi_{2}(\nu).$
The action of the Hamiltonian on  $\Psi(\nu)$ is given by:
$$ \hat{H}_{0} \Psi(\nu) = -{3}({2\gamma^2 \Delta_s \ell_p^2})^{-1}\left( \Psi(\nu + 2\lambda) - 2 \Psi(\nu) + \Psi(\nu - 2 \lambda) \right),
$$
with $\lambda=2 \sqrt{\Delta_{s}} \ell_p$. The key property is 
${\braket{\Psi_{1}(\nu),\Psi_{2}(\nu + 2\lambda)} 
=  \braket{\Psi_{1}(\nu_{} - 2 \lambda),\Psi_{2}(\nu_{})}}$ 
where $\nu_{}$ is in the support of both $\Psi_{1}(\nu)$ and $\Psi_{2}(\nu)$.
This is the statement of the unitarity of the shift operators $\braket{{e^{-i 2 \lambda b}} \Psi_{1}, \Psi_{2}} = \braket{\Psi_{1}, {e^{i2 \lambda b}}\Psi_{2}}$.
The symmetric nature of the shift operators appearing in $H_0$ implies the result.
} acting on $\sH_{lqg}$ 
\begin{equation}\label{eq89b}
{ {H}_{\Delta} \equiv  \frac{3 V_0}{8\pi G \gamma^2} \frac{1}{ {\Delta} \ell_p^2 } {{\sin^2\left( {\Delta^{\frac12}} \ell_p \, b\right)}},}
\end{equation}
which coincides with  \eqref{eq40}  to leading (zero) order in $\ell_p^2$. 
From \eqref{cosmoco} we obtain an operator associated to the (here dynamical) cosmological constant, namely
\be\label{lambdas}
{ {\Lambda}_{\Delta} \equiv\frac{3}{\gamma^2}  \frac{{{\sin^2\left( {\Delta^{\frac12}} \ell_p \, b\right)}}}{ {\Delta} \ell_p^2 } .}
\ee
In the pure gravity case, the cosmological constant is positive definite and bounded from above by the maximum value $\lambda_{\rm max}=1/(\gamma^{2}\ell_{p}^{2}\Delta )$. Negative cosmological constant solutions are possible when matter is added (see Appendix \ref{ultima}). The states (\ref{eige}) with $k=k_\Delta\equiv\sqrt{\Delta} \ell_p$ diagonalize the Hamiltonian, i.e.  \be \label{edeg}
{H}_{\Delta} \triangleright \ket{b_0; \Gamma^{\epsilon}_\Delta}=E_\Delta(b_0)\ket{b_0; \Gamma^{\epsilon}_\Delta},\ee with energy eigenvalues
\be\label{energy}
E_\Delta(b_0)=\frac{3 V_0}{8\pi G \gamma^2} \frac{1}{{\Delta} \ell_p^2 } {{\sin^2\left( {\Delta^{\frac12}} \ell_p \, b_0\right)}}.
\ee
States $\ket{b_0; \Gamma^{\epsilon}_\Delta}$ are also eigenstates of the cosmological constant with eigenvalue $\lambda_{\Delta}(b_0)=(8\pi G) E_\Delta(b_0)/V_0$. Notice that the energy eigenvalues do not depend on $\epsilon\in[0,4\sqrt{\Delta}\ell_p)$. Thus, the energy levels are infinitely degenerate with energy eigenspaces that are non-separable. This is not something peculiar of our model but a general property of the non-standard representation of the canonical commutation relations used in loop quantum cosmology. 

\subsection{On the interpretation of the $\epsilon$-sectors.}\label{epep}

It is customary in the loop quantum cosmology literature to restrict to a fixed value of $\epsilon$ in concrete cosmological models, as the dynamical evolution does not mix different $\epsilon$ sectors. The terminology {\em `superselected sectors'} is used in a loose way in discussions. However, these sectors are not superselected in the strict sense of the term because they are not preserved by the action of all the possible observables in the model, i.e. there are non trivial Dirac observables mapping states from one sector to another. The explicit construction of such observables might be very involved in general (as it is the usual case with Dirac observables); nevertheless, it is possible to exhibit them directly at least in one simple situation: the pure gravity case. In that case the shift operator (\ref{shifty}) with shift parameter $\delta$ commutes with the pure gravity Hamiltonian (the Hamiltonian constraint if we were in standard loop quantum cosmology) and maps the $\epsilon$ sector to the $\epsilon-4\delta$ sector. The analogous Dirac observables in a generic matter model can be formally described with techniques of the type used for the definition of evolving constants of motion \cite{PhysRevD.43.442}. No matter how complicated this might be in practise, the point is well illustrated by our explicit example in the matter free case \footnote{This point was independently communicated to us in the context of Dirac observables for isotropic LQC with a free matter scalar field \cite{madha}.}.  

Thus, different $\epsilon$ sectors are not superselected and therefore the infinite degeneracy of the energy eigenvalues of the Hamiltonian (which again we exhibit explicitly in the previous discussion only in the vacuum case) must be taken at face value. How can we understand this large degeneracy from the fact that there would be only a two fold one (associated with a contracting or expanding universe) if we had quantized the model using the standard Schroedinger representation or, in other words, the standard Wheeler-DeWitt quantization? The answer is to be found, we claim, in the notion of coarse graining: low energy observers only distinguish a two fold degeneracy for energy (or cosmological constant) eigenvalues: one the universe has a given cosmological constant, and two it is expanding or contracting. These are by the way the quantum numbers in the Wheeler-DeWitt quantization which plays the role in our context of the low energy effective quantum field theory formulation.  Such coarse observers are declared to be insensitive to the huge additional degeneracy of energy eigenstates encoded in the quantum number $\epsilon$. All these infinitely many states in the quantum cosmology representation must be considered as equivalent up to the two-fold degeneracy mentioned above.

In what follows, and for concreteness, we will consider combinations of states with two different values for $\epsilon$ only, i.e. on two different lattices. The idea of the previous paragraph will naturally produce a notion of coarse graining entropy associated to the intrinsic statistical uncertainty due to the inability for a low energy agent to distinguish these microscopically orthogonal states. Arbitrary superposition with $N$ different $\epsilon$-sectors would lead to similar results (the entropy capacity growing with the usual $\log(N)$). The $N=2$ case treated here makes some explicit calculations straightforward.

\subsection{Matter couplings and a model capturing its essential features.}\label{mama}

Here we discuss two simple matter models in order isolate the generic features of the influence of matter. At the end of the section we will define a simple and trivially solvable model capturing these features.
 
Perhaps the simplest matter model that would serve our purposes is the minimal and isotropic coupling to a Dirac Fermion defined in \cite{deBerredoPeixoto:2012xd}. After symmetry reduction the action  for matter is
\begin{equation}\label{eq136}
    S_{\rm F}(\eta,\bar \eta) = V_{0} \int_{\mathbb{R}} d\tau  \Big{[}\frac{i}{2} a(\tau)^{3} \left( \bar{\eta} \gamma^{0} \dot{\eta} - \dot{\bar{\eta}} \gamma^{0} \eta \right) - m N(\tau) a(\tau)^{3} \bar{\eta} \eta \Big{]},
\end{equation}
from which we read the Fermionic contribution to the Hamiltonian 
\ba\label{eq137}
    {H}_{\rm F} &=& m  N(\tau) a^{3}(\tau) \bar{\eta} \eta=-m N(\tau) p_\eta\gamma_0 \eta\n \\
&=&\frac{m}{a^3}p_\eta\gamma_0 \eta,
\ea
where $(\Pi,\Psi)$ are the Fermionic canonical variables $\Pi\equiv  (V_0a)^{3/2} \psi^{\dagger} $ and $\Psi\equiv (V_0a)^{3/2}  \psi$ \cite{Thiemann:2007zz}, and in the second line we use the unimodular condition $N=V_0^{-1}a^{-3}$. In the quantum theory the non trivial anticommutator is  $\{\eta, p_{\eta}\}=\mathbf{1}$ with the rest equal to zero. This is achieved by writing $\eta=\sum_s \left(a_s u^s e^{-i m t}+b^\dagger_s v^s e^{i m t}\right)$ with non trivial anti-commutation relations for the creation and annihilation opetrators $\{a_r,a^\dagger_s\}=\delta_{rs}=\{b_r,b^\dagger_s\}$, and $u^s e^{-i m t}$ and $v_s e^{i m t}$ a complete basis of solutions of the Dirac equation for positive and negative frequency respectively \cite{Peskin:1995ev}. In our model we can have either the vacuum state, one or two Fermions which saturates the Pauli exclusion principle. If we assume normal ordering the contribution to the unimodular energy is
\be
{H}_{\rm F} =\frac{m n}{a^3}.
\ee  
where $n=0,1,2$ is the occupation number for the Fermion. If instead of the condition $N=V_0^{-1}a^{-3}$ we had used $N=1$ (where time is comoving time) then the energy contribution would have been just $m$ for which we have a clear physical intuition: a single fermion homogeneously distributed in the universe contributes to the Hamiltonian with its total mass. The factor $1/a^3$ in the previous expression comes from the unimodular condition.

 In the case of Wheeler-de-Witt quantization the contribution of the fermion becomes singular at the big-bang $a=0$. In loop quantum cosmology such a quantity remains bounded above due to loop quantum gravity discreteness. Indeed, using the inverse volume quantization given in reference \cite{Ashtekar:2011ni}  one has 
\be\label{hF}
\Hat{H}_{\rm F} \triangleright \ket{\psi} 
= - m \sum_\nu \ket{\nu}  h_{\rm F}(\nu; \sqrt{\Delta} \ell_p) \Psi(\nu,\eta) ,
\ee
where
\be
h_{\rm F}(\nu; \lambda)\equiv \frac{1}{4 \lambda^2}{\left(|\nu +2 \lambda |^{\frac 12}-|\nu -2 \lambda|^{\frac 12}\right)^2}.
\ee
We notice that $h_{\rm F}(\nu; \sqrt{\Delta} \ell_p)< 1$ and decays like $1/\nu$ for $\nu\to \infty$\footnote{There is a great degree of ambiguity in writing the inverse volume operators. Perhaps the simplest is the one introduced in \cite{WilsonEwing:2012bx} that we will actually used in the concrete computations of the Appendix \ref{scalarf}. For more discussion on this see \cite{Singh:2013ava} and references therein.}.
One could in principle add this term to the free Hamiltonian and solve the unimodular time independent Schroedinger equation
\be
(\Hat{H}_{0}+\Hat H_{\rm F} )\triangleright \ket{\psi}=E \ket{\psi}. 
\ee
Solutions can be interpreted in the sense of scattering theory starting with free wave packets for large $\nu$ picked around some value of the cosmological constant \eqref{lambdas} or energy \eqref{energy}. 

The case of a the coupling with a scalar field is formally very similar, specially in the simplified case where we assume it to be massless. Following \cite{Ashtekar:2011ni} and using the unimodular condition $N=V_0^{-1}a^{-3}$ we get
\be\label{scaly}
H_{\phi}=-\frac{p_{\phi}^2}{8\pi^2 \gamma^2 \ell_p^4 \nu^2}.
\ee  
This leads to
\be\label{hf}
\Hat{H}_{\rm \phi} \triangleright\ket{\psi}
= -m \sum_\nu  \ket{\nu}  h_{\rm \phi}(\nu; \sqrt{\Delta} \ell_p) \Psi(\nu,\phi) ,
\ee
where
\be
h_{\rm \phi}(\nu; \lambda)\equiv \frac{p_{\phi}^2}{16 \lambda^4}{\left(|\nu +2 \lambda |^{\frac 12}-|\nu -2 \lambda|^{\frac 12}\right)^4}.
\ee
The momentum $p_{\phi}$ commutes with the Hamiltonian and thus it is a constant of motion. If we consider an eigenstate of $p_{\phi}$ then the problem reduces again to a scattering problem with a potential decaying like $1/\nu^2$ when we consider solving the time independent Schroedinger equation  
\be
(\Hat{H}_{0}+\Hat H_{\rm \phi}) \triangleright  \ket{\psi}=E  \ket{\psi}. 
\ee

Therefore, both the Fermion as well as the scalar field models (which are closer to a possibly realistic scenario) seem tractable with a slight generalization of the standard scattering theory to the discrete loop quantum cosmology setting.  However, the main objective in this section is to illustrate an idea in terms of a concrete and simple toy model. With this idea in mind we will modify the structure suggested by the Fermion coupling and the scalar field coupling and simply add an interaction term where the `long distance interaction' term represented by the function $F(\nu; \lambda)$ is replaced by a short range analog $F(\nu; \lambda)\propto \delta_{\nu,0}$. The qualitative properties of the scattering will be the same and the model becomes sufficiently trivial for straightforward analytic computations. The results for the more realistic free scalar field model have been dealt with numerically and are presented in the Appendix \ref{scalarf}.

 \begin{figure}[h] \centerline{\hspace{0.5cm} \(
\begin{array}{c}
\includegraphics[width=8cm]{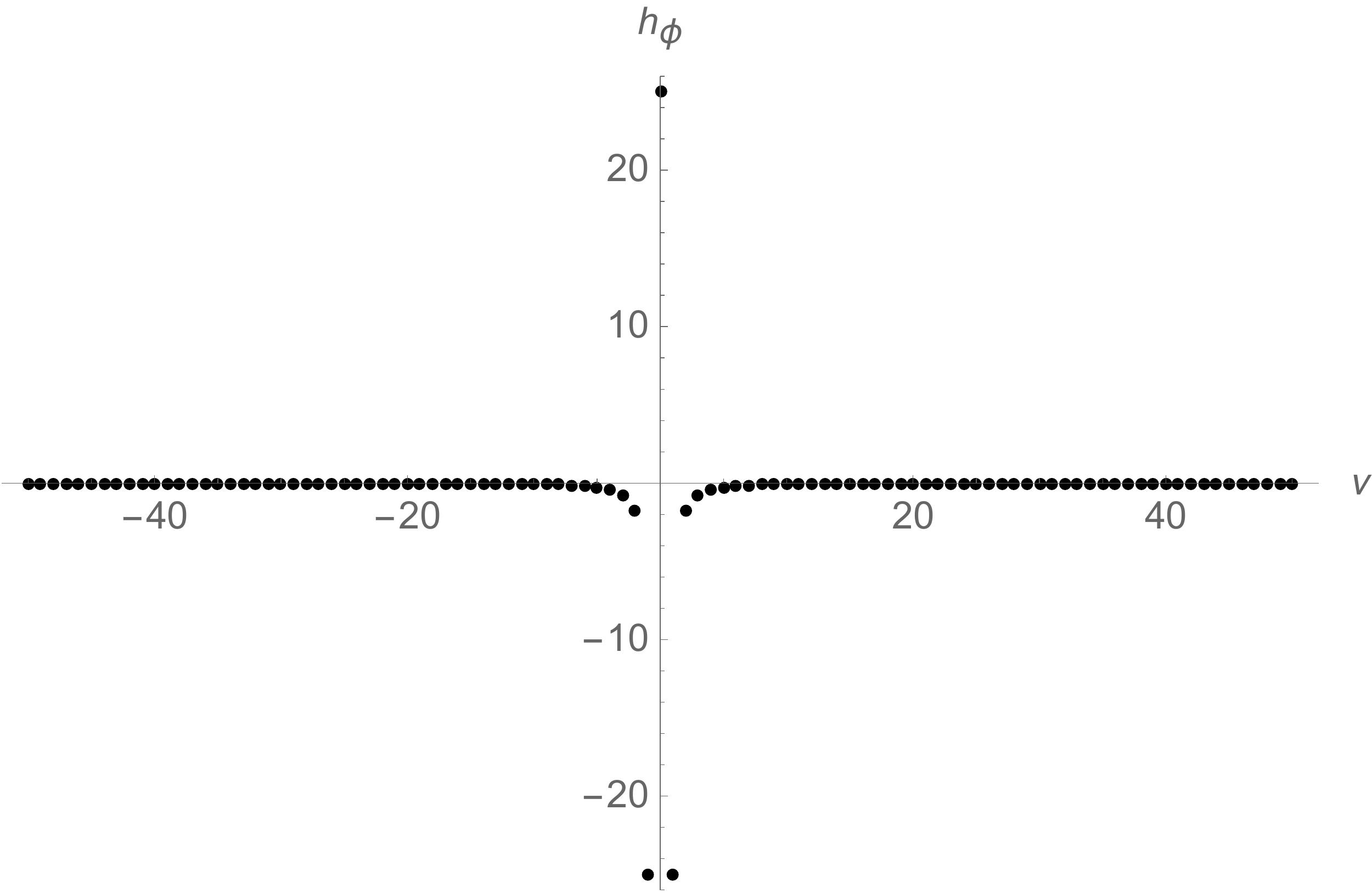}
\end{array}  \ \ \ \ \ \ \ \ \ \ \ \begin{array}{c}
\includegraphics[width=8cm]{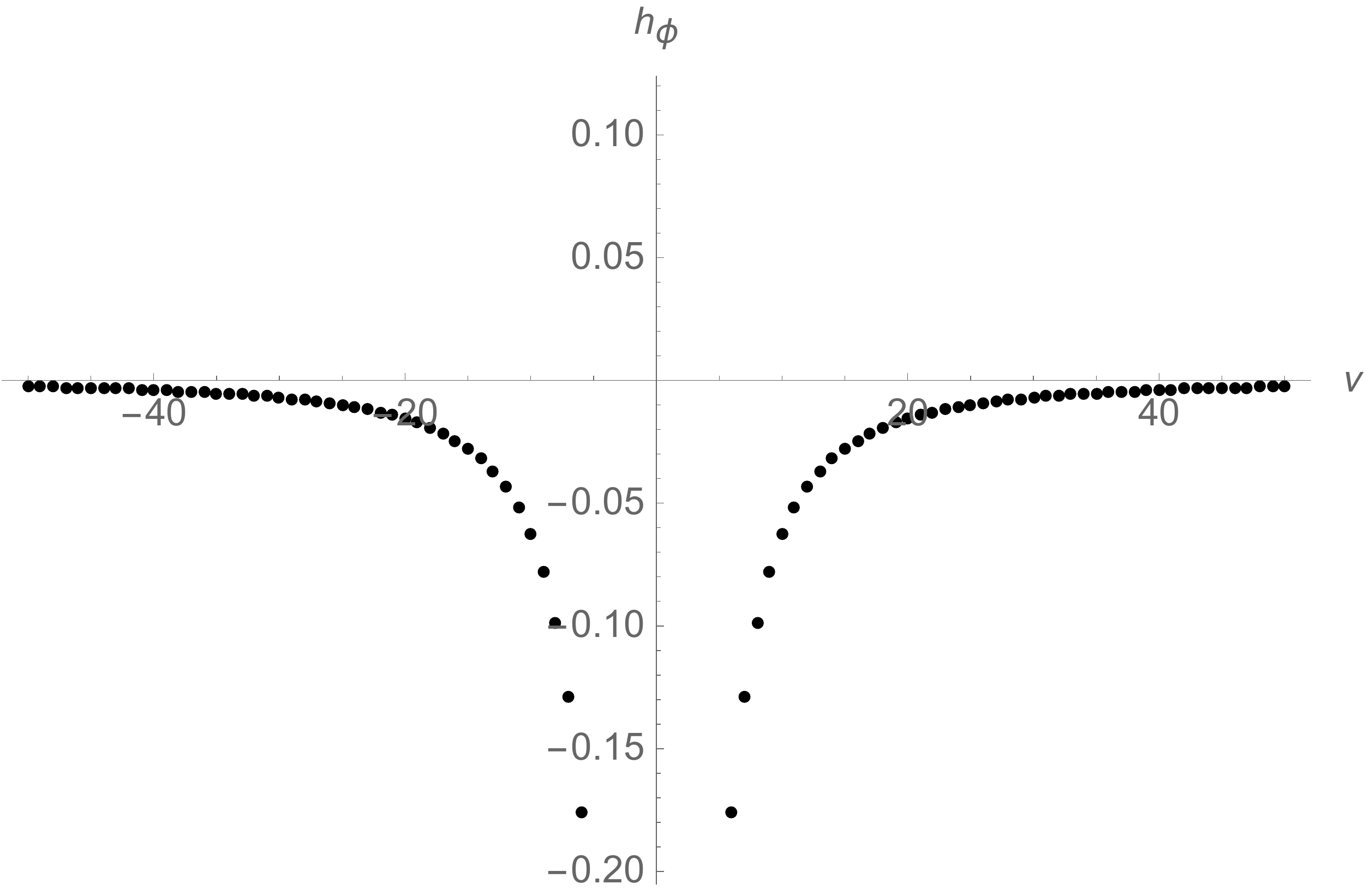}
\end{array} 
\)}
\caption{The function $h_{\rm \phi}(\nu; \lambda)$ evaluated on an epsilon sector containing $\nu=0$ for $p_{\phi}=10$ in natural units and $\lambda=1/2$ is plotted using two different ranges. On the left we see that the function is finite near $\nu=0$. On the right we can see that it behaves like $-\nu^{-2}$ for large values of $\nu$. This fucntion can be seen as the effective potential where an asymptotically free state of the universe (pure gravity with cosmological constant state or asymptotically DeSitter state) scatters. If the cosmological constant is negative there are bound states whose superposition can be used to define semiclassical universes oscillating in an endless series of big-bangs and big-crunches (see Section \ref{ultima}).}
\label{potential}
\end{figure}

For that we consider an interaction that kicks in at $\nu=0$:
\begin{equation}\label{eq147}
    \Hat{H} = \Hat{H}_{0} + \mu \Hat{H}_{\rm int},
\end{equation}
where $\mu$ is a dimensionless coupling, $\Hat{H}_{0}$ is given in \eqref{eq149}, and 
$\Hat{H}_{\rm int}$ is
\be\label{eqq36}
\Hat{H}_{\rm int} \triangleright \ket{\psi}\equiv \sum_\nu  \left( \ell^{-4}_p \frac{V_0}{\sqrt{\Delta}} \right)\ket{\nu} \frac{\delta_{\nu,0}}{\sqrt{\Delta}} \Psi(0).
\ee
We have added by hand an interaction Hamiltonian that switches  on only when the universe evolves through the {\em would-be-singularity} at the zero volume state. The key feature of the $\Hat{H}_{\rm int}$ is that---as its more realistic relatives matter Hamiltonians (\ref{hF}) and (\ref{hf})---it breaks translational invariance and thus it leads to different dynamical evolution for different $\epsilon$-sectors. 

\subsection{Solutions as a scattering problem}\label{sca}

The scattering problem is very similar to the standard one in one-dimensional quantum mechanics; however, one needs to take into account the existence of the peculiar degeneracy of energy eigenvalues contained in the $\epsilon$ sectors; see Sections \ref{qqq} and \ref{epep}.
We will consider, for simplicity, the superposition of only two states supported on two lattices respectively: the lattice $\Gamma_\Delta^\epsilon$  with $\epsilon=0$ for the first one and the one with $\epsilon=2\sqrt{\Delta} \ell_p$ for the second one. The degenerate eigenstates of the shift operators (\ref{eige}) with eigenvalues $\exp(i2kb)$ will be denoted
\ba\label{dedito}
\ket{b,1}\equiv \ket{b;\Gamma_\Delta^0},\ \ \ \ 
{\rm and} 
\ \ \ \ 
\ket{b,2}\equiv \ket{b;\Gamma_\Delta^{2\sqrt{\Delta} \ell_p}},
\ea
respectively, while we will denote by $\Gamma^1$ and $\Gamma^2$ the corresponding underlying lattices. The immediate observation is that states supported on $\Gamma^2$ (superpositions of $\ket{b,2}$) will propagate freely because they are supported on a lattice that does not contain the point $\nu=0$ where the interaction is non trivial. On the other hand, states supported on $\Gamma^1$ (superpositions of $\ket{b,1}$)  will be affected by the interaction at the big bang.
Before and after the big bang the universe's evolution of the second type of states is well described by the eigenstates of the Hamiltonian (\ref{eq149}) described in Section \ref{qqq}.  Such asymmetry of the interaction on different $\epsilon$-sectors is not an artifact of the simplicity of the interaction Hamiltonian. This is just a consequence of the necessary breaking of the shift invariance for any realistic matter interaction as we argued in the previous section and we show explicitly in Appendix \ref{scalarf} (see Figure \ref{potential}). 

Therefore, the non trivial scattering problem concerns only states on the lattice $\Gamma^1= \{\nu= 2n\sqrt{\Delta} \ell_p \ | \ n \in \mathbb{Z} \}$ that is preserved by the Hamiltonian and contains the point $\nu=0$. In order to solve the scattering problem  
we consider an in-state of the form
\begin{equation}\label{eq151}
\ket{\psi_k}  =
\ket{\nu}\begin{cases}
e^{- i \frac k 2\nu} + A(k) \, e^{i \frac{k}{2} \nu} & \text{($\nu \ge 0$)} \\
B(k)\, e^{- i \frac k2 \nu} & \text{($\nu \le 0$)},
\end{cases}.
\end{equation}
where $\nu\in \Gamma^1$,  and $A(k)$ and $B(k)$ are coefficients depending on $k$.  For suitable coefficients, such states are eigenstates of the full Hamiltonian \eqref{eq147}.
Arbitrary solutions (wave packets) can then be constructed in terms of appropriate superpositions of these `plane-wave' states.
   
 We can compute the scattering coefficients $A(k)$ and $B(k)$ from the discrete (finite difference) time-independent Schrodinger equation
\ba\label{tise}
 \left(\Hat{H}_{0}+\Hat{H}_{\rm int} \right)\triangleright \ket{\psi} &=& E \ket{\psi}\ea
which amounts to the following finite difference equation in the $\nu$ basis:   
 \ba\label{shishi}
\n \sum_\nu \left(-\frac{3 V_0}{8\pi G \gamma^2} \frac{1 }{2\Delta \ell_p^2}  \Big{[}\Psi(\nu - 4\sqrt{\Delta} \ell_p) + \Psi(\nu + 4\sqrt{\Delta} \ell_p)  - 2\Psi(\nu) \Big{]} + \frac{V_0 \mu}{{\Delta} \ell_p^4} \delta_{\nu,0} \Psi(0) - E(k)\,  \Psi(\nu)\right)  \ket{\nu}=0.
\ea
The matching conditions on $\nu = 0$ are given by:
\begin{align}\label{eq100}
    \begin{split}
         1 + A(k) = B(k)& \\
       -\frac{3}{16\pi G\gamma^2 \Delta \ell_p^2}  \Big{[}\Psi(-4 \sqrt{\Delta} \ell_p) + \Psi(4
        \sqrt{\Delta} \ell_p) -2 \Psi(0) \Big{]} + &  \frac{\mu}{{\Delta} \ell_p^4} \, \Psi(0)  = \frac{E(k)}{V_0} \Psi(0),
    \end{split}
\end{align}
where the first equation comes from continuity at $\nu=0$, the second equation from the time independent Shroedinger equation. The solution of the previous equations is
\ba \label{eq152}
    A(k) &=& \n \frac{- i \Theta(k)}{1 + i \Theta(k)} \\
     B(k) &=& \frac{1}{1 + i \Theta(k)}.\ea
where    
\ba      \label{61-s}
     \Theta(k) \equiv  \frac{16\pi \gamma^2} {3} \frac{\mu}{\sin(2 k \sqrt{\Delta} \ell_p)}.
\ea
We consider an ${\rm in}$-state of the form (valid for early times)
\be\label{42}
\ket{\psi_{in}, t}= \frac{\pi}{\sqrt{ 2\Delta} \ell_p} \int db \left( \psi(b; b_0,\nu_0)\ket{b,1}  + \psi(b; b_0,\nu_0) \ket{b,2}\right) e^{-i E_{\Delta}(b) t },
\ee
where $\psi(b; b_0,\nu_0)$ is a wave function picked at some $b=b_0$ value and $\nu=\nu_0$.  Notice that we are superimposing two wave packets supported on lattices $\Gamma^1$ and $\Gamma^2$ respectively.  We can now write the pure in-density matrix
\ba\label{inicial}
&&  \rho_{\text{in}}(t) = \frac{\pi^2}{{ 2\Delta} \ell^2_p} \int db\ db^\prime \,   e^{i [E_{\Delta}(b)- E_{\Delta}(b')] t}  \\ && \ \ \ \ \ \ \ \ \  \ \ \ \ \ \ \times\n    \Big{[} \ket{b', 1} \psi(b^\prime; b_0,\nu_0)  +\ket{b', 2}\psi(b^\prime; b_0,\nu_0) \Big{]} {\Big [}  \bra{b^{},1} \overline \psi(b; b_0,\nu_0)  + \bra{b^{}, 2}\overline \psi(b; b_0,\nu_0) {\Big ]} .
\ea 
which scatters  into the  out-density matrix 
\ba\label{den}
&&  \rho^{}_{\text{out}}(t) = \frac{\pi^2}{{ \Delta} \ell^2_p} \int db\ db^\prime \,  e^{i [E_{\Delta}(b)- E_{\Delta}(b')] t}    \Big{[}\bra{b^{},1}  \overline \psi(-b; b_0,\nu_0) \overline A(-b) + \bra{b^{},1}  \overline \psi(b; b_0,\nu_0)  \overline B(b)+\bra{b^{},2} \overline \psi(b; b_0,\nu_0) \Big{]} \n  \\ && \Big{[} \ket{b', 1}  \psi_{}(-b^\prime) e^{-i b^\prime \overline \nu} A(-b^\prime) +  \ket{b', 1} \psi_{}(b^\prime) e^{i b^\prime \overline \nu} B(b^\prime)+ \ket{b', 2} \psi(b^\prime; b_0,\nu_0) \Big{]} \n.
\ea
Let us assume that $\psi(b)$ is highly picked at some $b_0$ so that we can substitute the integration variables $b$ and $b^\prime$ by $b_0$ and have a finite dimensional representation of the reduced density matrix after the scattering (this step is rather formal, it involves an approximation but it helps visualising the result).
In the relevant $4\times4$ sector (with basis elements ordered as $\left\{\ket{1,b_0},\ket{1,-b_0},\ket{2,b_0}, \ket{2,-b_0}\right\}$) we get the matrix representation
\be
{\bm \rho}_{in}=\left(
\begin{array}{cccc}
 \frac{1}{2} & 0 & \frac{1}{2} & 0 \\
 0 & 0 & 0 & 0 \\
 \frac{1}{2} & 0 & \frac{1}{2} & 0 \\
 0 & 0 & 0 & 0 \\
\end{array}
\right) \ \ \ \rightarrow \ \ \ 
{\bm \rho}_{out}=\frac{1}{2}\left(\begin{matrix}  |B(b_0)|^2 & \overline A(-b_0) B(b_0) & B(b_0)  & 0 \\
 A(-b_0) \overline{B(b_0)} & |A(-b_0)|^2  & A(-b_0) & 0\\
 \overline{B(b_0)}&  \overline {A(-b_0)}& 1 & 0 \\
 0& 0& 0& 0 \end{matrix}\right).
\ee

\subsection{Matter coupling produces a coarse-graining entropy jump at the big-bang} \label{sisi}

A reduced density matrix encoding the notion of coarse graining associated with the low energy equivalence of the $\epsilon$-sectors is defined by tracing over the discrete degree of freedom labelling the component of the state in either the $\Gamma_1$ or the $\Gamma_2$ lattices. In other words, tracing over the two (macroscopically indistinguishable) $\epsilon$-sectors, namely \be 
\bra{b}\rho^{\rm \va R}\ket{b^{\prime}}\equiv \sum_{i=1}^2\bra{b,i}\rho\ket{b^\prime,i}.\ee
In other words, the subspace of the Hilbert space we are working with is the one supported on two different epsilon sectors $\sH(\Gamma^1)\oplus \sH(\Gamma^2)\subset \sH_{\rm lqc}$ which, as the two terms are isomorphic $\sH(\Gamma^1)\approx \sH(\Gamma^2)\approx \sH_0$,   $\sH(\Gamma^1)\oplus \sH(\Gamma^2)\subset \sH_{\rm lqc}$  can be written as
\be
\sH_0\otimes \C^2\subset \sH_{\rm lqc}.
\ee  
The coarse graining is defined by tracing over the $\C^2$ factor. 
This implies that from the previous $4\times 4$ matrix we obtain the $2\times 2$ reduced density matrices.
The reduced density matrix ${\bm \rho}^{\va \rm R}_{in}$ remains pure, explicitly 
\be
{\bm \rho}^{\va \rm R}_{in}=\frac{1}{2}\left(\begin{matrix} 1& 1  \\
 1 &  1 \end{matrix}\right).
\ee
 Nevertheless, the reduced density matrix ${\bm \rho}^{\va \rm R}_{out}$ is now mixed, namely
\be
{\bm \rho}^{\va \rm R}_{out}=\frac{1}{2}\left(\begin{matrix}  1+ |B(b_0)|^2 & \overline A(-b_0) B(b_0)  \\
 A(-b_0) \overline{B(b_0)} &  |A(-b_0)|^2\end{matrix}\right).
\ee
We can now compute the entanglement entropy. To first order in the cosmological constant the result is
\be
\delta S=\log (2)-\frac{3 \Delta  \Lambda 
   {\ell_p}^2}{128 \pi ^2 \gamma ^2 \mu ^2}+\sO(\Lambda^2 \ell_p^4)\ee
The behaviour as a function of $b$ is shown in Figure \ref{Entro-Pietro}. In Appendix \ref{cge} we discuss an alternative definition of coarse graining with the same qualitative implications.

 \begin{figure}[h] \centerline{\hspace{0.5cm} \(
\begin{array}{c}
\includegraphics[width=10cm]{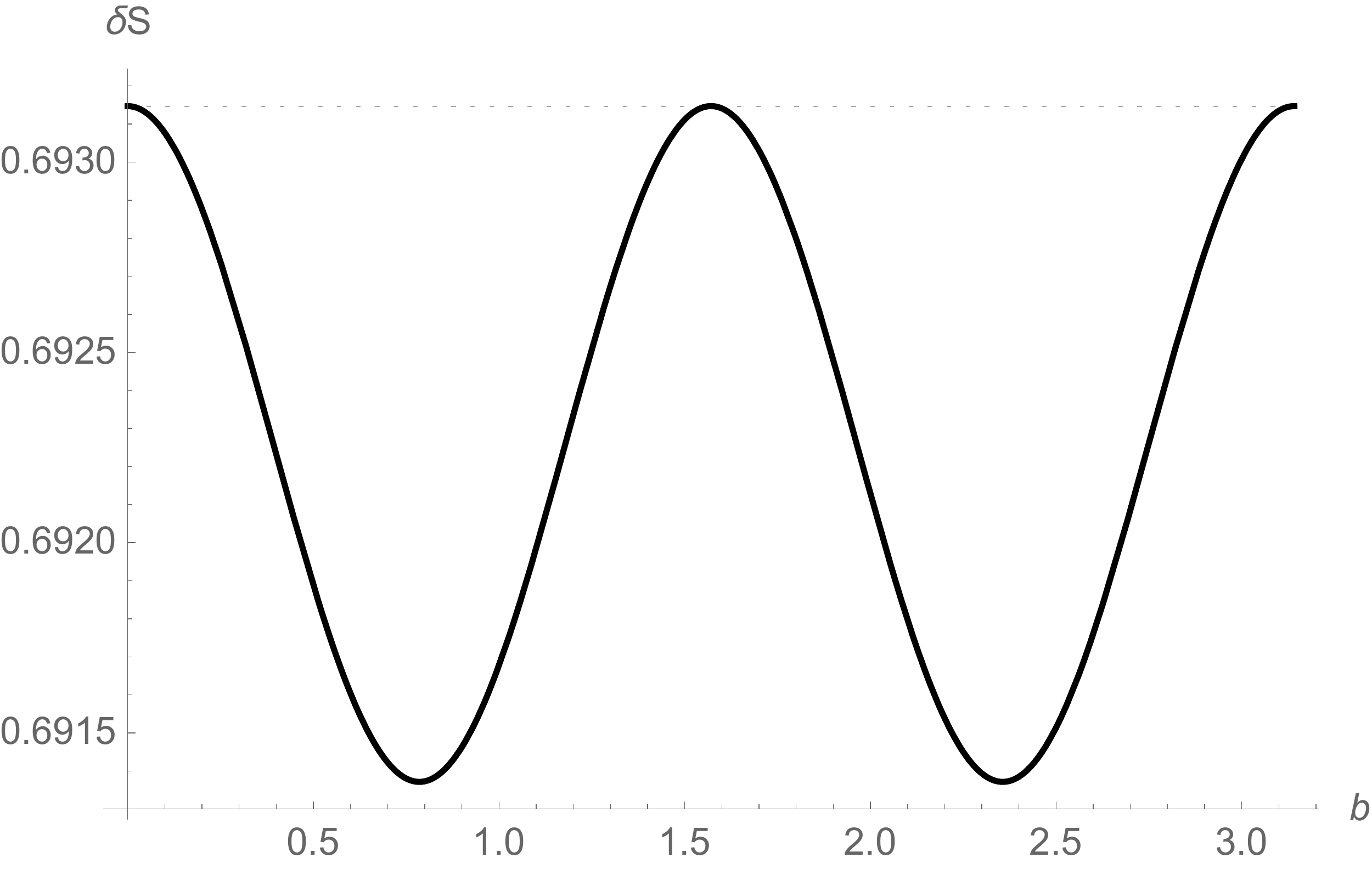}
\end{array}  
\)}
\caption{The curve represented by a thin line is the entropy jump $\delta S$ as a function of $b$ in Planck units for $\gamma=\mu=\Delta=1$.
The small $b\ell_p$ behaviour in \eqref{bbb} is apparent. The entropy is periodic for $b\ell_p\in [0,\pi]$ as expected from \eqref{61}. The dotted line represents the maximum possible entropy which is $\log[2]$ in our model. }
\label{Entro-Pietro}
\end{figure}

\section{Quantum cosmology on a superposition of backgrounds.}\label{maco}

In the first part of the paper we have seen how the fact that the Hilbert space of loop quantum cosmology is vastly larger 
than the standard Schroedinger representation  implies (via coarse graining) that the coarse graining
entropy would go up generically through the evolution across the big-bang {\em would-be-singularity}.
In this section we explore another closely related feature that leads to an apparent non-unitary evolution 
when dynamics is probed by a low energy agent. This new key property of the fundamental quantum dynamics in loop quantum cosmology is tight to the fact that the Hamiltonian defining evolution can only be defined if an area-gap $\Delta$ is provided (see Section \ref{qqq}). The quantization of the Hamiltonian reviewed in \ref{qqq} needs an input from a UV background structure.   We will see here that the loop quantum cosmology model can be extended naturally to admit superpositions of such microscopic structures and that 
such extension generically leads to the dynamical development of correlations  between the macroscopic and the microscopic degrees of freedom.  If the microscopic degrees of freedom are assumed to remain hidden to low energy observers, then such correlations lead to an apparent violation of unitarity in the low energy description where pure states evolve into mixed states.

\subsection{The UV input in quantum cosmology: revisiting the $\overline \mu$ scheme. }\label{sfour}

The $\bar \mu$ scheme was designed to avoid an inconsistency of an early model of loop quantum cosmology with the low energy limit (or large universe limit) of loop quantum cosmology \cite{Green:2004mi}. The problem arises from the effective compatification of the connection variable $c$ due to the polymer regularization of the Hamiltonian with a fixed fiducial scale $\mu$ which implies that $c$ and $c+4\pi/\mu$ are dynamically identified. This leads to anomalous deviations from classical behaviour  in situations where the variable $c$ is classically expected to be unbounded for large universes. This can be seen clearly in the present situation where the unimodular Hamiltonian (\ref{eq40}) is given, in $(c,p)$ variables, by 
\begin{equation}\label{contrast}
    H=\frac{3 V_0}{8\pi G} \frac{c^2}{\gamma^2 |p|}. 
\end{equation}
For non-vanishing energies (or equivalently non-vanishing cosmological constant) the conservation of the Hamiltonian implies that $c$ grows as $|p|\propto a^2$, i.e., $c$ grows without limits as the universe expands so that no matter how small $\mu$ is, anomalous effects due to the compactification of the $c$ become relevant at macroscopic scales \cite{Noui:2004gy}. As no quantum gravity effects seem acceptable in the large universe regime for a model with finitely many degrees of freedom, this anomaly is seen as an inconsistency of the model.

The $\bar \mu$ scheme solves this inconsistency by `renormalizing' the regulating scale $\mu$ as the universe grows (recall equation \eqref{eq17}). The interesting thing is that such renormalization is justified by quantum geometry arguments that link the mini superspace model of loop quantum cosmology to the geometry of a microscopic background state in the full theory. The argument uses explicitly the idea that the low energy degrees of freedom (dynamical variable of loop quantum cosmology) arise from the coarse graining of the fundamental ones in loop quantum gravity. 
 
Here we review the construction of the $\bar \mu$ scheme as described in \cite{Ashtekar:2011ni}. Consider a fundamental quantum geometry state $\ket{s}$ in the Hilbert space of loop quantum gravity,  representing a microscopic state on top of which the quantum cosmological coarse grained dynamics will eventually be defined. Such underlying fundamental state will have to be approximately homogeneous and isotropic up to some scale $L>\ell_p$ with respect to the preferred foliation defining the co-moving FLRW observers at low energies. If that is the case then such space slices can be divided into (approximately) cubic 3-cells of physical side length $L$ which all have approximately equivalent quantum geometries. The area of a face of such cubic cells in Planck units will be denoted $\Delta_s$ so that $
L^2=\ell_{p}^{2}\Delta_s $. Note that $\Delta_s$ is a property of the underlying microstate: an area eigenstate if the microstate is an eigenstate, or an area expectation value if the state is sufficiently peaked on a quantum geometry and has small fluctuations around it. A simple realization is the one where  $\Delta_s$ is an area eigenvalue, and the important assumption is that $\Delta_s$ is the same for all cells (this encodes the homogeneity of the microsocopic state). 
Consider the area of a large two dimensional surface (the face of a fiducial cell $\cal V$) whose area is measured by the low energy (coarse grained) quantity $p$ used as configuration variable in loop quantum cosmology. We naturally would expect that $|p|\gg \ell_p^2$ or alternatively that     
\begin{equation}\label{eq75}
    N \ell_{p}^{2}\Delta_s  = |p|, \end{equation}
   where $N$ denotes the number of microscopic cells contained in the coarse grained surface (a face of $\cal V$), and  $N\gg 1$.  The fiducial cell has fiducial coordinate volume $V_0$ and hence fiducial side coordinate length $V_0^{{1}/{3}}$. Therefore, the fiducial coordinate length $\bar \mu$ of the microscopic homogeneity cells is given by the relation 
\begin{equation}\label{eq76}
    N (\bar{\mu} V_{0}^{1/3})^{2} = V_{0}^{2/3}.
\end{equation}
Combining the previous two equations one recovers equation (\ref{eq17}), namely
\be\label{mumy}
\bar \mu^2_s\equiv \bar{\mu}^2=\frac{\ell_{p}^{2}\Delta_s }{|p|},
\ee
i.e., the fiducial scale $\bar \mu$ is dynamical: as the universe grows (and $|p|$ becomes large), the underlying fiducial length scale decreases. The fiducial regularization scale (\ref{eq76}) depends on the fundamental state $\ket{s}$ via the quantity $\Delta_s$, hence we denote it $\bar\mu_s$. When such dynamical scale is used in the regularization of the quantum cosmology Hamiltonian the effective compactification scale for $c$ grows like $|p|$ and the inconsistency previously discussed is avoided. This is transparent in terms of the new canonical pair $(b, v)$.  From equation (\ref{eq38}) we have that $b= c\bar \mu_s/(\sqrt{\Delta_s} \ell_p)$, in contrast with $c$ (see \eqref{contrast}),  remains constant (see \eqref{eq40}) in the De Sitter universe. The quantization of the Hamiltonian presented in Section \ref{qqq} introduces an effective compatification of the variable $b$ whose dynamical effect is now only relevant when the cosmological constant  approaches one in Planck units. This can be seen from \eqref{eq89b}. The cosmological constant is bounded from above by its natural value in Planck units due to the underlying quantum geometry structure while the anomalous IR behaviour is avoided (the problems exhibited in the model studied in \cite{Green:2004mi} are also resolved).   

The previous is the standard account of the motivation of the $\bar \mu$ scheme of \cite{Ashtekar:2006wn} with the little twist (which is very important for us here) that $\Delta_s$ need not be the lowest area eigenvalue of loop quantum gravity. In the usual argument the microscopic state is thought to be built from a special homogeneous spin network (geometry eigenstate) with all spins equal to the fundamental representation. This implies that, in the above construction, $\Delta_s=\Delta_{1/2} \equiv 2\pi \gamma\sqrt{3}$. The observation here is that $\Delta_s$ can take different values according to the microscopic properties of the underlying quantum geometry state. One could take for instance all spins equal to the vector representation and then have $\Delta_s=\Delta_{1} \equiv 4\pi \gamma\sqrt{2}$ instead, or take $j$ arbitrary and use $\Delta_s=\Delta_j$.  It is important to point out that such possibility can arise naturally in quantum cosmology models obtained in the group field theory framework \cite{Gielen:2013kla, Oriti:2016qtz, Oriti:2016ueo}.

As we have seen in Section \ref{sfour}, the field strength regularization, and hence the Hamiltonian, depend on the value $\Delta_s$ of the background (approximately homogeneous) spin network state $\ket{s}$ through the dynamical scale $\bar \mu_s$.  In this way, the dynamics of loop quantum cosmology establishes correlations with the a microscopic degree of freedom in the underlying loop quantum gravity fundamental state. As such degree of freedom (the area eigenvalue $\Delta_s$ of the minimal homogeneity cells) is quantum, it is natural to model the system by a tensor product Hilbert space $\sH \equiv \sH_m\otimes \sH_{\rm lqc}$ where $\sH_m$ is the Hilbert space representing the microscopic degree of freedom encoded in the minimal homogeneous cell operator (whose eigenvalues we denote $\Delta_s$), and $\sH_{\rm lqc}$ the standard kinematical Hilbert space of loop quantum cosmology. 

General states in $\sH$ can be expressed as linear combinations of product states $\ket{s}\otimes \psi$ in the respective factor Hilbert spaces. The quantum Hamiltonian has a natural definition on such states and therefore on the whole of $\sH$, namely
\begin{equation}\label{eq146}
    \Hat{H} \triangleright \left(\ket{s} \otimes \psi\right) = \ket{s} \otimes \ \Hat{H}_{\Delta_{s}} \triangleright\psi,
\end{equation}
where $\Hat{H}_{\Delta_{s}}$ is the usual loop quantum cosmology  Hamiltonian in the $\bar\mu_s$ scheme, which in our particular case is defined in equation \eqref{eq89b} with regulator $\Delta=\Delta_s$. 

Notice that the previous extension of the standard loop quantum cosmology framework  to the larger Hilbert space $\sH$ is also natural from the perspective of the full theory. Indeed the generally accepted regularization procedure of the Hamiltonian constraint in loop quantum gravity (first introduced by Thiemann  \cite{Thiemann:1996aw} and further developped in recent analysis---see Varadarajan and Ladda \cite{Laddha:2017fdi} and references therein) is state dependent in that the loops defining the regulated curvature of the connection are added on specific nodes of the state where the Hamiltonian is acting upon.
This feature finds its analog in the action \eqref{eq146} where the regulating scale $\Delta_s$ depends on the state $\ket{s}\in \sH_m$.

In order to simplify the following discussion we will restrict states in $\sH_m$ even further and consider  a subspace $\mathfrak{h}=\C^2\subset \sH_m$, i.e. we will model the situation where the underlying microscopic state is an arbitrary superposition of only two fixed microscopic homogeneous spin-network states. For example we take  \be 
\mathfrak{h}\equiv {\rm span} \Big{[}\ket{+}, \ket{-}\Big{]}, \ee
where $\ket{\pm}\in \sH_m$ are two suitable orthogonal  background states (these two states will be conveniently picked below).   
From the infinitely dimensional Hilbert space $\sH_m$ we are  now selecting a single {\em q-bit} subspace $\C^2$. The Hilbert space of our model is 
\be\label{hilbert}
\sH=\mathfrak{h}\otimes \sH_{\rm lqc}.
\ee
The factor $\mathfrak{h}$ represents additional microscopic (hidden to low energy observers) UV degrees of freedom, while $\sH_{\rm lqc}$ encodes the data that under suitable circumstances (e.g. when the universe is large) represent the low energy cosmological degrees of freedom. 

In this way we see that in addition to the intrinsic degeneracy of energy eigenvalues analyzed in the first part of this paper, there is another candidate for microscopic degree of freedom associated to the regularization of the Hamiltonian action via the $\bar \mu$-scheme. Both mechanisms are proper of the present loop quantum cosmology toy model but reflect generic properties of the full theory of loop quantum gravity. More generally, we expect similar features to be present in any quantum gravity approach where smooth geometry is only emergent from a discrete fundamental theory. 

From now on we adopt the convenient notation $\ket{s}$ with $s=\pm$ for such prefered basis elements of $\mathfrak{h}$. 
With this notation, and using \eqref{eq89b}, the Hamiltonian \eqref{eq146} becomes \begin{equation}\label{eq149}
\begin{split}
 \Hat{H}_{0}\triangleright\left(\ket{s} \otimes \ket{\psi}\right) &=\frac{3 V_0}{8\pi G \gamma^2}    \frac{1 }{\Delta_s \ell_p^2} \left( {{\sin( \sqrt{\Delta_s} \ell_p \, b)}} \right)^{2} \triangleright\ket{s} \otimes  \ket{\psi} \\
    &= -\frac{3 V_0}{8\pi G \gamma^2}   \sum_\nu \frac{1}{2 \Delta_s \ell_p^2} \ket{s} \otimes \ket{\nu} \Big{[}\Psi(\nu - 4\sqrt{\Delta_s} \ell_p) + \Psi(\nu + 4\sqrt{\Delta_s} \ell_p)  - 2\Psi(\nu) \Big{]},
\end{split}
\end{equation}
where $\Psi(\nu)\equiv \braket{\nu|\psi}$.
 \begin{figure}[h] \centerline{\hspace{0.5cm} \(
 \begin{array}{c}
\includegraphics[width=11cm]{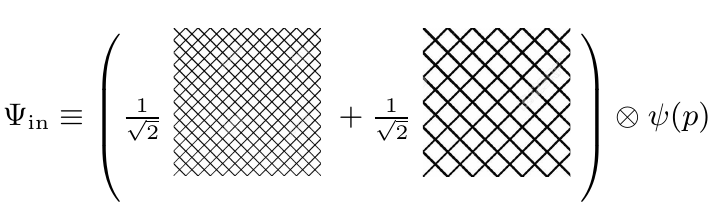}
\end{array}
\)}
\caption{Schematic representation of the state of interest. There are two different UV structures with dynamical implications via the $\bar \mu$-scheme. The state represented here has trivial correlations with the microscopic structure and would lead to a zero initial entanglement entropy state as defined by the reduced density matrix where the background state is traced out.}
\label{structu}
\end{figure}
 
Now, the only special feature of the basis $\ket{\pm}$ is that it is preferred from the perspective of the regularization of the effective (unimodular) loop quantum cosmology Hamiltonian. Consequently, a natural question to the quantum theory is how the dynamics would look if the initial state is arbitrary in the factor $\mathfrak{h}$? More precisely, what if we consider the linear combination of two background spin networks $\frac{1}{\sqrt{2}}(\ket{+} +\ket{-}) \in \mathfrak{h}$ times some loop quantum cosmology wave function as depicted in Figure  \ref{structu}? To answer this question we consider a special initial state where correlations between the low energy and the UV degrees of freedom are not present. This will lead to a reduced density matrix---tracing out the microscopic space $\mathfrak{h}$ in \eqref{hilbert}---that is {\em pure} initially, the form of such state is illustrated in Figure \ref{structu}. At the same time we want to be able to diagonalize the Hamiltonian with such uncorrelated initial  states; more precisely this boils down to diagonalizing both $H_{\Delta_+}$ and $H_{\Delta_-}$ in $\sH_{\rm lqc}$. This implies that the factor $\psi(\nu)\in \sH_{\rm lqc}$, in Figure \ref{structu}, must be supported on a lattice $\Gamma^{\epsilon}_\Delta$ that is left invariant by the action of both $H_{\Delta_+}$ and $H_{\Delta_-}$ (left invariant in the sense that the shift operators in the definition of the Hamiltonian only relate points of $\Gamma^{\epsilon}_\Delta$ and never map points out).  This can be achieved by assuming that $\sqrt{\Delta_+}=m \sqrt{\Delta_-} $ for some natural number $m$.  For simplicity we will take $m=2$ from now on \footnote{One might be worried that is hard to achieve if one sticks to the form of the area spectrum of loop quantum gravity. This is however simply a model and the link with the full theory (remember) must be taken at the heuristic level. Nevertheless,  solutions do exist for instance $m=4$ for $j_+=3$ and $j_-=1/2$.}.  The parameter $\epsilon$ will be taken so that the lattice $\Gamma^{\epsilon}_\Delta$ contains the point $\nu=0$. This is a standard choice. With all this the invariant lattice, denoted $\Gamma_{\Delta_-}$, is
\be \Gamma_{\Delta_-}\equiv \Gamma^{\epsilon=0}_{k=2\sqrt \Delta_-\ell_p}.\ee 
Note that in the notation described below (\ref{dedito}) we have that $\Gamma_{\Delta_-}=\Gamma_1\cup \Gamma_2$.

%

The choices made above are not mandatory. One could have chosen a different initial state. The previous choice is particularly interesting here because it would lead to a reduced initial density matrix that is pure and hence and initially vanishing entanglement entropy. Other states would involve correlations and would therefore carry a non vanishing entropy load from the beginning. For the discussion that interests us here and for the analogy with black hole evaporation it is more transparent to set the entropy to zero initially. 

\begin{figure}[h] \centerline{\hspace{0.5cm} \(
\begin{array}{c}
\includegraphics[width=11cm]{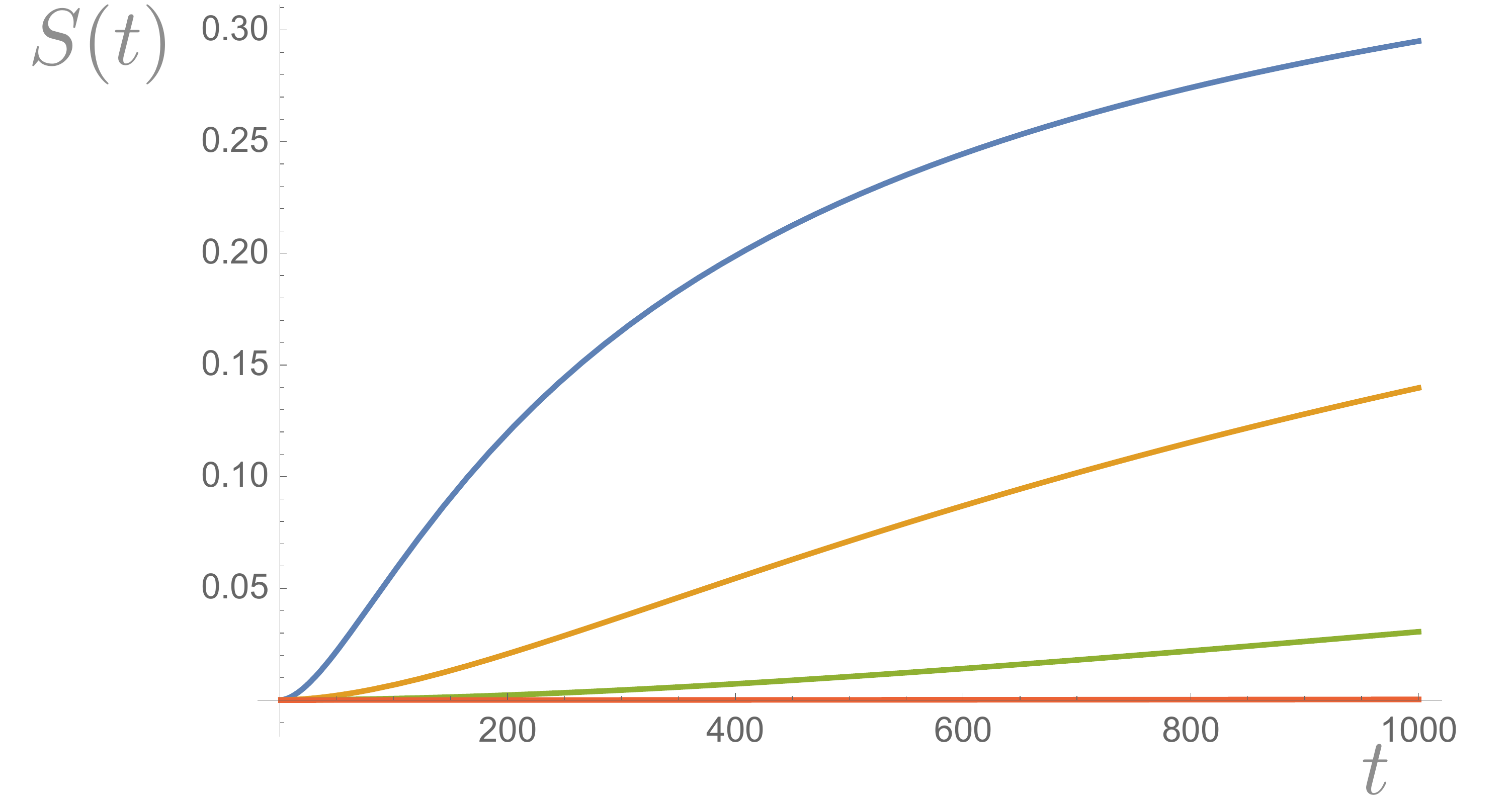}
\end{array}
\)}
\caption{ Here we plot $S(t)$ as a function of time for a gaussian wave packet centred at $b=2.5\  10^{-2}$,  $b=5 \ 10^{-2}$,  $b=7\  10^{-2}$, and $b=10^{-1}$ with width $\sigma=b$ respectively. Numerical integration plus the approximation (\ref{eq158}) was used with the assumption $2 (\Delta_+-\Delta_-)/\gamma^2=1$, all in Planck units. As $b$ grows the scalar curvature (the cosmological constant) grows and the rate at which entropy increases grows as well. For $b\ll 1$ an effective unitary evolution is recovered.}
\label{fifi}
\end{figure}

An arbitrary  (unimodular) loop quantum cosmology state  associated to such choice of background state can be expressed as: 
\ba\label{eq153}
\n    {\Psi_{\text{in}}(\nu,t)} &=&\bra{\nu} \frac{1}{\sqrt{2}} \sum_{s}  \ket{s} \otimes |\Psi_{\rm in}(t)\rangle  \\  &=& \frac{1}{\sqrt{2}} \sum_{s}  \ket{s} \otimes \Big{[}\delta_{\Gamma_{\Delta_-}}(\nu) \int\limits_{0}^{\frac{\pi}{\sqrt{\Delta_-} \ell_p}}dk \ \psi(k;b_0,\nu_0) \exp\large(- i E_{s}(k) t  \Large{)} \Big{]} 
\ea
where $\psi(k;b_0,\nu_0)$ is a properly normalized function peaked at $k= b_0$ and $\nu=\nu_0$. 
The initial state in the momentum representation is given by:
\ba \label{eq156}
    {\Psi_{\text{in}}(b,t)}&=&\sum_{\nu\in \Gamma^0_{\Delta_-}}  \braket{b,1\cup 2|\nu}\braket{\nu |\Psi_{\rm in}(t)}\\
%
 &=& \frac{\pi }{\sqrt{\Delta_{-}} \ell_p} \sum_{s} \ket{s} \otimes \psi(b; b_0,\nu_0)  e^{- i E_{s}(b) t}  \n
\ea
where in the first line we used the natural extension of the notation introduced in \eqref{dedito} where $\ket{b,1\cup 2}$ means an eigenstate of the corresponding shift operators (\ref{eige}) supported on the lattice $\Gamma_{\Delta_-}=\Gamma_1\cup\Gamma_2$.  Notice that we can also write
\be
\ket{b,1\cup 2}=\ket{b,1}+\ket{b, 2},
\ee keeping in mind that terms on the {\em r.h.s.} are individually eigenstates of the shift operators with twice the lattice spacing of $\Gamma_{\Delta_-}$. We also used \ba
&& \sum_{\nu\in \Gamma^0_{\Delta_-}}  \exp\left(i\frac{b-k}{2} \nu\right)=\frac{\pi }{\sqrt{\Delta_-} \ell_p}  \delta(b-k). 
\ea
%
We can write then 
\ba \label{keykey}
    {\Psi_{\text{in}}(t)= \sum_{s}   \int {\rm D}b \ket{s} \otimes \ket{b,1\cup 2} \psi(b; b_0,\nu_0)  e^{- i E_{s}(b) t} } ,
\ea
where
\be{\rm D}b\equiv \frac{\pi}{\sqrt{2 \Delta_{-}}\ell_p} db,\ee
is the Haar measure on the circle of circumference ${\pi}/{\sqrt{2 \Delta_{-}}\ell_p}$. We notice from  \eqref{keykey}  that even when our initial state contains no correlations between the low energy degrees of freedom represented by $b$ and the microscopic degrees of freedom encoded in $\ket{s}$ at $t=0$, quantum correlations between the two will develop with time due to the non trivial dependence of the energy spectrum with $s$.   
Even when this is quite clear from \eqref{keykey} one can state this fact in an equivalent way by analysing the (pure) density matrix $\rho_{\rm in}(t)\equiv\ket{\Psi_{\rm in}(t)} \bra{\Psi_{\rm in}(t)}$, whose matrix elements in the $b$ basis are:
\ba\label{eq157}
 \rho_{\rm in}(t)&\equiv&  \sum_{s,s^\prime}  \int {\rm D} b\ {\rm D} b^\prime \left({\overline \psi(b; b_0,\nu_0)  \psi(b'; b_0,\nu_0)  \ e^{i [E_{s}(b) - E_{s'}(b')]t}}\right) \n \\  &\times&  \ket{b', 1\cup 2}\ket{s'} \bra{b^{}, 1\cup 2}\bra{s^{}}.
\ea 
As coarse grained observers are assumed to be insensitive to the microscopic structure that is here encoded in the `spin' quantum number $s$, low energy physical information is encoded in the reduced density matrix \ba 
\rho_{\rm in R}(t)&\equiv&  \sum_{s}   \int {\rm D} b\ {\rm D} b^\prime   \left(  {\overline \psi(b; b_0,\nu_0)\psi(b'; b_0,\nu_0)  \ e^{i [E_{s}(b) - E_{s}(b')]t}}\right) \\  &\times&  \ket{b', 1\cup 2}\ket{s'} \bra{b^{}, 1\cup 2}\bra{s^{}},
 \ea  
 which can be simply be written as
\be\label{rere}
\rho_{\rm in R}(t)=\frac{1}{2}\sum_s \ket{\Psi_s(t)}\bra{\Psi_s(t)}.
\ee 
where
\be
\ket{\Psi_s(t)}\equiv \int  {\rm D}b\ {\psi(b; b_0,\nu_0)  \ e^{-i E_{s}(b)t}} \,  \ket{b, 1\cup 2}
\ee
 Notice that \eqref{rere} is only pure at $t=0$ and becomes mixed due to the correlations evoked above as time passes. One can compute the entanglement entropy $S(t)\equiv -{\rm Tr}\left[\rho_{\rm in R}(t)\log(\rho_{\rm in R}(t)) \right]$ which turns out to be given by the simple analytic expression (see Appendix \ref{entropia})
 \ba\label{rentro}
 S(t)={-\log\left(1-\frac{\delta}{2} \right)-\frac{\delta}{2}\log\left(\frac{\delta}{1-\frac{\delta}{2}}\right)},
 \ea 
 where 
 \be\label{pito}
 \delta(t)\equiv 1-\left|\int {\rm D}b \, \overline \psi(b; b_0,\nu_0) \psi(b; b_0,\nu_0)  e^{i [E_{+}(b) - E_{-}(b)]t}\right|. 
 \ee
 For generic wave packets $\psi_s(b)$ the entanglement entropy is a monotonic growing function of time which grows asymptotically to the maximally mixed situation $S_{\max}=\log(2)$ (see an example in Figure \ref{fifi}).
 
A more intuitive picture can be obtained from a suitable expansion of the energy eigenvalues \eqref{energy}  in powers of the label $b\ell_p$
\begin{equation}\label{eq158}
   E_s(b)= \frac{3 V_0}{8\pi G \gamma^2}   \frac{1}{ \Delta_s \ell_p^2} \left( {\sin( \sqrt{\Delta_s} \ell_p b )} \right)^{2}= \frac{3 V_0}{8\pi G \gamma^2}   b^2-\frac{V_0}{8\pi G \gamma^2}  \Delta_s \ell_p^2 b^4+b^2\sO(\ell_p^4 b^4).
\end{equation}
Such expansion makes sense in that 
it allows for the identification of the low energy effective Hamiltonian (the one that one would define in a purely Wheeler-DeWitt quantization) plus corrections that involve interactions with the underlying discrete structure of LQG here represented by the spin $s$ degree of freedom. Namely, we can read from the previous expansion
\be
H_{\rm eff}\equiv H^0_{\rm eff}(b)+ {\Delta H} (b, s),
\ee 
where $\hat H^0_{\rm eff}(\hat b)\equiv \frac{6}{\gamma^2} \hat b^2$ is the Wheeler-DeWitt Hamiltonian and the additional term an interaction with the 
environment represented by the underlying discrete structure represented by the dependence on $S_1$ (a hidden degree of freedom from the low energy continuum perspective). Of course the hats in the previous equation denote operators in a different representation (the continuum Schroedinger representation)  that is not unitarily equivalent to the `fundamental' polymer representation introduced in Section \ref{qqq} and used in the LQC setup (recall for instance that the operator $\hat b$ does not even exist in the polymer representation).

The lack of purity for $t>0$ of the reduced density matrix \eqref{rere} is due to correlations that develop between the low energy degree of freedom $b$ and the hidden microscopic degree of freedom $s$ via this non trivial interaction Hamiltonian.   This means that generically (i.e. for arbitrary initial states $\psi_s(b)$) the fundamental evolution would seem to violate unitarity, from the perspective of low energy observers, due to the decoherence with the microscopic quantum geometric structure. Notice however that for states $\psi_s(b)$ picked at sufficiently small $\bar k$, i.e., $\bar k \sqrt{\Delta_s} \ell_p\ll 1$,  we have from \eqref{lambdas} that 
\be \label{lambdasa}
\Lambda(b)\approx {3\gamma^{-2}} b^2\ee and the density matrix \eqref{rere} is pure for all times. 

More precisely, we can translate the criterion for the absence of decoherence with the underlying microscopic discrete structure in terms of the value of the cosmological constant of the given state. For an eigenstate of the Hamiltonian the relation is given by $\Lambda\equiv E_s(b)$. Therefore the criterion for the absence of decoherence in terms of the cosmological constant is
\be\label{deco}
\Delta_s \ell_p^2 \gamma^2 \Lambda\approx \ell^2_p \Lambda \ll 1
\ee
 Interestingly for states with low values of the cosmological constant in natural units---equivalent semi-classically to the scalar curvature $R$ in our matter free model---define a decoherence free subspace. When the cosmological constant does not satisfy the condition (\ref{deco}) decoherence with the microscopic structure is turned on and maximized for $\Lambda$ of order one in Planck units: notice incidentally that due to the polymer quantization the cosmological constant is bounded by \be \Lambda_{\rm max} = \frac{3}{\gamma^2 \Delta_{\frac{1}{2}} \ell_p^2}.\ee For low values of $\Lambda$ unitarity is recovered in the effective description that ignores the microscopic structure.
 
Decoherence takes place here due to an interaction between the low energy coarse degrees of freedom and the microscopic discreteness in the underlying quantum geometry background but in way (in our simple model) that the energy and (hence the cosmological constant) is conserved. However, the presence of decoherence suggest the possibility for a natural deviation of this idealized absence of dissipation: generically decoherence and dissipation often come together.    
Therefore, a surprising and unexpected  consequence of our analysis is the suggestion of a natural channel for the  relaxation of a large cosmological constant due to the possibility of dissipative effects associated to the decoherence pointed out here.

Incidentally, all this shows it is only in the limit of low values of $E$ (small cosmological constant) that the coarse graining that leads from the full theory of loop quantum gravity to the minisuperspace description of loop quantum cosmology is well defined. This is not surprising and only confirms the usual intuition that drives the construction of models of loop quantum cosmology. However, it opens the door for a qualitative understanding of the necessity of decoherence effects in more general situations. For instance, the standard $\bar \mu_s$ construction suggests that coarse graining is weaker at the big bang where the Hamiltonian evolution \eqref{eq149} takes the universe through  the $\nu=0$ states. During this high (spacial) curvature phase it is natural to expect that the higher corrections in \eqref{eq158} (describing the interaction with the microscopic Planckian structure) can no longer be neglected.   

Interestingly, there is another way to make decoherence go away. This is due to the asymptotic behaviour of the separation of area eigenvalues in loop quantum gravity  which imply that for large $\Delta_s$ there are states such that $\Delta_s-\Delta_{s^\prime}\approx \Delta_s \exp(-\pi \sqrt{2\Delta_s/3})$  \cite{Barbero:2017jxa}. Therefore, in the continuum limit $\Delta_s-\Delta_{s^\prime}\ll 1$ the dynamical entanglement growth of our model can be made as small as wanted.

\subsection{Matter coupling produces entanglement entropy jump at the big-bang}\label{soso}

In the pure gravity case we can make decoherence be as small as wanted by choosing  states with a cosmological constant that is sufficiently small. Here we show that this is no longer possible once matter is added and that there is a generic development of correlations with the UV degrees of freedom in the evolution across the {\em would-be-singularity}: an initially pure state (reduced low energy density matrix) evolves generically into a mixed state (reduced low energy density matrix) after the big-bang.  

In order to see this in more detail 
we just need to write the matter Hamiltonians acting in the Hilbert space \eqref{hilbert}. 
One needs the natural generalization of the expressions written in Section \ref{mama} to the present context.
For instance for the scalar field coupling equation \eqref{hf} becomes 
\be\label{hhff}
\Hat{H}_{\rm \phi} \triangleright\left( \ket{s} \otimes \ket{\psi} \right)
= -m \sum_{\nu\in \Gamma} \ket{s} \otimes \ket{\nu}  h_{\rm \phi}(\nu; \sqrt{\Delta_s} \ell_p) \Psi(\nu,\phi) ,
\ee
where
\be\label{77}
h_{\rm \phi}(\nu; \lambda)\equiv \frac{p_{\phi}^2}{16 \lambda^4}{\left(|\nu +2 \lambda |^{\frac 12}-|\nu -2 \lambda|^{\frac 12}\right)^4}.
\ee
The momentum $p_{\phi}$ commutes with the Hamiltonian and thus is its a constant of motion. As before,  if we consider an eigenstate of $p_{\phi}$ then the problem reduces again to a scattering problem with a potential decaying like $1/\nu^2$ when solving the time independent Schroedinger equation  
\be
\Hat{H}_{0}+\Hat H_{\rm \phi} \triangleright\left(\ket{s} \otimes \ket{\psi}\right)=E \left(\ket{s} \otimes \ket{\psi}\right). 
\ee

From the discussion of \ref{mama} we can capture the basic qualitative effect of matter interaction by considering a simple solvable model where the matter contribution is concentrated at a single event a the big-bang. Non of the qualitative conclusions that follow depend on this simplification, and the more realistic free scalar field model can be dealt with (some results are shown in Appendix \ref{scalarf}). With some extra effort one could actually analyze the a more realistic model (say the one defined by \eqref{hhff}) but the conclusion will remain the same.
Therefore we consider 
\begin{equation}\label{eq147b}
    \Hat{H} = \Hat{H}_{0} + \mu \Hat{H}_{\rm int},
\end{equation}
where $\mu$ is a dimensionless coupling, $\Hat{H}_{0}$ is given in \eqref{eq149}, and 
$\Hat{H}_{\rm int}$ is the generalization of \eqref{eqq36}
\be
\Hat{H}_{\rm int} \triangleright\left( \ket{s} \otimes \ket{\psi} \right)\equiv \sum_\nu\hat{O}\ket{s}\otimes \ket{\nu} \frac{\delta_{\nu,0}}{\sqrt{\Delta_s}} \Psi(0)
\ee
where $\hat O$ is a self adjoint operator in $\mathfrak{h}=\C^2$.  A natural and simple model for this operator is to choose \be \hat O\equiv \ell^{-4}_p \frac{V_0}{\sqrt{\Delta_s}}.\ee
This choice is formulated in the notation introduced below \eqref{eq149} and inspired by the analogy with a spin system. We have added by hand an interaction Hamiltonian that switches  on only when the universe evolves through the {\em would-be-singularity} at the zero volume state. This encodes the idea of the intrinsic uncertainty of the peculiar construction of the mini-superspace model of loop quantum cosmology that we discussed in Section \ref{sfour}. The discrete local degrees of freedom must be important close to the big bang and symmetry reduction must fail in some way that can only be correctly described if a full quantum gravity theory is available. Here we model such unknown dynamics 
in the simplest fashion available to us here,  which consists of including the possibility for the background state $\ket{s}$ (representing in spirit the underlying quantum geometry) to be modified by the dynamics via $\Hat{H}_{\rm int} $.

Here we proceed as in Section \ref{sca} while keeping in mind that, in the present case, there are two distinct cases at hand given by the two possible values $\Delta_{\pm}$.  Let us consider an in-state of the form
\begin{equation}\label{eq151}
\ket{k,s}  =
\ket{s} \otimes \ket{\nu}\begin{cases}
e^{- i \frac k 2\nu} + A_s(k) \, e^{i \frac{k}{2} \nu} & \text{($\nu \ge 0$)} \\
B_s(k)\, e^{- i \frac k2 \nu} & \text{($\nu \le 0$)},
\end{cases}.
\end{equation}
where $A_s(k)$ and $B_s(k)$ are coefficients depending on $k$ and (in contrast with the case in Section \ref{sca}) now also on $s=\pm 1$ (with $\ket{\pm}$ the eigenstates of $\hat S_z$).  For suitable coefficients, such states are eigenstates of the Hamiltonian $H_0$ as well as the full Hamiltonian \eqref{eq147}.
Arbitrary solutions (wave packets) can then be constructed in terms of appropriate superpositions of these `plane-wave' states.

\ba \label{coco}
    A_s(k) &=& \n \frac{- i \Theta_s(k)}{1 + i \Theta_s(k)} \\
     B_s(k) &=&  \frac{1}{1 + i \Theta_s(k)}.\ea
where    
\ba      \label{61}
     \Theta_s(k) \equiv  \frac{16\pi \gamma^2} {3} \frac{\mu}{\sin(2 k \sqrt{\Delta_s} \ell_p)}.
\ea
One can superimpose the previous eigenstates to produce wave packets (semiclassical states) for the wave function of the universe that are picked at some value $\nu_0$ of the rescaled volume (see footnote \ref{footy}). Wave packets will evolve in time according to the Schroedinger equation which in our case is just a discrete analog of the one corresponding to a free particle in quantum mechanics with an interaction term at the `origin' $\nu=0$. If we start with a state that is sufficiently picked around $\nu_0$ for $\nu\gg \ell_p$  initially, then the state can be described in terms of the supperposition \eqref{keykey} where the explicit values of the coefficients $A_s(b)$ and $B_s(b)$ does not appear. Equation \eqref{42} is generalized to 
\ba \label{inicial}
    {\Psi_{\text{in}}(t\ll 0)}
 &=&   \int {\rm D}b \left( \ket{b,1} \psi(b; b_0,\nu_0)+\ket{b,2} \psi(b; b_0,\nu_0) \right) e^{-i E_{-}(b) t }  \\ \n  &+&  \int {\rm D}b \left( \ket{b,1} \psi(b; b_0,\nu_0)+\ket{b,2} \psi(b; b_0,\nu_0) \right) e^{-i E_{+}(b) t }    . 
\ea
The coefficients \eqref{coco} enter the expression of the scattered wave packet at late times which becomes
\ba \label{final}
 &&    {\Psi_{\text{out}}(t\gg 0)}
 = \\ \n &&   \int {\rm D}b\  \ket{-} \otimes  \ket{b, 1\cup 2}  \Big{[}  \psi(-b; b_0,\nu_0)  A_-(-b) + \psi(b; b_0,\nu_0)  B_-(b)\Big{]}  e^{- i E_{-}(b) t} 
 +\\ &&    \int {\rm D}b\   \ket{+} \otimes  \Big{[}\ket{b,1} \left(\psi(-b; b_0,\nu_0)  A_+(-b) + \psi(b; b_0,\nu_0) B_+(b)\right)+ \ket{b,2} \psi_{}(b; b_0,\nu_0) \Big{]}  e^{- i E_{+}(b) t} \n  .
\ea
Note that the solution of the scattering problem for the $E_+(b)$ eigenvalues is asymmetric with respect to the components of the in state supported on $\Gamma_1$ and $\Gamma_2$. Indeed the states $\ket{b,2}$ are eigenstates of the Hamiltonian directly because they are not supported on  $\nu=0$ and hence they do not `see' the interaction: this is capture by trivial scattering coefficients for this component. 

\subsection{Entropy associated with the entanglement with the UV degrees of freedom}

From the previous initial state we can calculate (by tracing over the factor $\frak h$, see \eqref{hilbert}) the initial reduced density matrix 
\ba\label{ini}
\rho^{\rm R}_{\text{in}}(t) 
&=& \int {\rm D}b\ {\rm D}b^\prime \,   e^{i [E_{+}(b)- E_{+}(b')] t}  \\ && \ \ \ \ \ \ \ \ \  \ \ \ \ \ \ \times\n    \Big{[}  \ket{b', 1} \psi (b^\prime; b_0,\nu_0)  +\ket{b', 2}\psi(b^\prime; b_0,\nu_0) \Big{]} {\Big [}  \bra{b^{},1} \overline \psi(b; b_0,\nu_0)  + \bra{b^{}, 2}\overline \psi(b; b_0,\nu_0) {\Big ]}  \n \\
&+& \int {\rm D}b\ {\rm D}b^\prime \,   e^{i [E_{-}(b)- E_{-}(b')] t}  \n \\ && \ \ \ \ \ \ \ \ \  \ \ \ \ \ \ \times\n    \Big{[} \ket{b', 1} \psi (b^\prime; b_0,\nu_0)  +\ket{b', 2}\psi(b^\prime; b_0,\nu_0) \Big{]} {\Big [}  \bra{b^{},1} \overline \psi(b; b_0,\nu_0)  + \bra{b^{}, 2}\overline \psi(b; b_0,\nu_0) {\Big ]}.
\ea 
The reduced density matrix after the big bang is
\ba\label{denfiR}
&&  \rho^{\rm R}_{\text{out}}(t) = \int {\rm D}b\ {\rm D}b^\prime \,  e^{i [E_{+}(b)- E_{+}(b')] t}  \\ && \Big{[}   \bra{b,1} \Big( \overline \psi(-b; b_0,\nu_0)  \overline A_+(-b) +  \overline \psi(b; b_0,\nu_0)  \overline B_+(b)\Big)+\bra{b,2}\overline \psi(b; b_0,\nu_0) \Big ]   \n \\ &&   \Big{[}\ket{b,1}\left( \psi(-b^\prime; b_0,\nu_0) A_+(-b^\prime) + \psi(b^\prime; b_0,\nu_0)  B_+(b^\prime)\right)+\ket{b,2} \psi(b^\prime; b_0,\nu_0) \Big{]}  + \n\\
&& e^{i [E_{-}(b)- E_{-}(b')] t}   \n  \Big{[} \ket{b^\prime,1\cup2}  \psi(-b^{\prime}; b_0,\nu_0)  A_-(-b^\prime) +  \ket{b^\prime,1\cup2}  \psi(b^{\prime}; b_0,\nu_0) B_-(b^\prime)\Big{]}   \\ && \times\Big[ \bra{b,1\cup2}  \overline\psi(-b; b_0,\nu_0) \overline  A_-(-b) +   \bra{b,1\cup2} \overline \psi(b; b_0,\nu_0)  \overline B_-(b)\Big]  
\n ,
\ea
where $\alpha_+=1/4$ and $\alpha_-=1$ and $\delta_{s+}$ is unity when $s=+$ and vanishes when $s=-$.   Then the non vanishing entries of the reduced density matrix are
\ba\label{66}
 &&  {\rho_{out}^{R \ 11}}(b_0, b_0)=\frac14\left( |B_+(b_0)|^2+|B_-(b_0)|^2\right)\n \\
 &&  {\rho_{out}^{22}}(b_0, b_0)=\frac14\left( 1 +|B_-(b_0)|^2\right)\n 
 \\
 &&  {\rho_{out}^{R\ 12}}(b_0, b_0)=\frac14\left( B_+(b_0) +|B_-(b_0)|^2\right)=\n \overline{{\rho_{out}^{R\ 21}}(b_0, b_0)}
 \\
 &&  {\rho_{out}^{R\ 11}}(-b_0, -b_0)=\frac14\left( |A_+(-b_0)|^2+|A_-(-b_0)|^2\right)\n 
  \\
  &&  {\rho_{out}^{R\ 22}}(-b_0, -b_0)=\frac14\left(|A_-(-b_0)|^2\right)\n 
 \\
 &&  {\rho_{out}^{R\ 12}}(-11b_0, -b_0)=\frac14\left(|A_-(-b_0)|^2\right)=\n \overline{{\rho_{out}^{R\ 21}}(-b_0, -b_0)}\\
 &&  {\rho_{out}^{R\ 11}}(b_0, -b_0)=\frac14\left( \overline A_+(-b_0) B_+(b_0) +\overline A_-(-b_0) B_-(b_0)\right)=\n \overline{{\rho_{out}^{R\ 11}}(-b_0, b_0)}\n \\
 &&  {\rho_{out}^{R\ 22}}(b_0, -b_0)=\frac14\left(\overline A_-(-b_0) B_-(b_0)\right)=\overline{{\rho_{out}^{R\ 22}}(-b_0, b_0)}\n \\
 &&  {\rho_{out}^{R\ 21}}(b_0, -b_0)=\frac14\left(\overline A_+(-b_0)+\overline A_-(-b_0) B_-(b_0) \right)=\n \overline{{\rho_{out}^{R\  12  }}(-b_0, b_0)}\\
 &&  {\rho_{out}^{R\ 12 }}(b_0, -b_0)=\frac14\left(\overline A_-(-b_0) B_-(b_0) \right)= \overline{{\rho_{out}^{R\ 21}}(-b_0, b_0)}.
\ea
The matrix $\rho_{out}^{R}$ is positive definite, ${\rm Tr}[\rho_{out}^{R}]=1$ and $\rho_{out}^{R}=\rho_{out}^{R \dagger}$. In the case $b_0 \ell_p\ll 1$  we have
\be
     \Theta_s(b_0) \approx  \frac{8\pi \gamma^2 \mu} {3} \frac{1}{ b_0 \sqrt{\Delta_s}\ell_p}.
\ee
We can now compute the entanglement entropy jump $\delta S$ to first leading order in $b_0 \ell_p/\mu$. The result (expressed in terms of the cosmological constant in this regime, namely  \eqref{lambdasa}) is 
\be\label{bbb}
\delta S=\delta_0 S-\frac{3
   \Delta_-   \ell_p^2 \log (3)}{128 \pi
   ^2 \gamma ^2 \mu ^2}\ \Lambda +\sO(\Lambda^2\ell_p^4),
\ee
where $\delta_0 S=2\log(2)-\frac34 \log(3)$.
The previous equation shows that the entropy jump is non trivial at crossing the big-bang {\em would-be-singularity},  even in the low cosmological (low energy limit) where 
(according to the analysis of the previous section) decoherence with the microscopic Planckian structure can be neglected during the time the universe is large. Information is unavoidably degraded (it seems lost for low energy observers) during the singularity crossing.

 \begin{figure}[h] \centerline{\hspace{0.5cm} \(
\begin{array}{c}
\includegraphics[width=10cm]{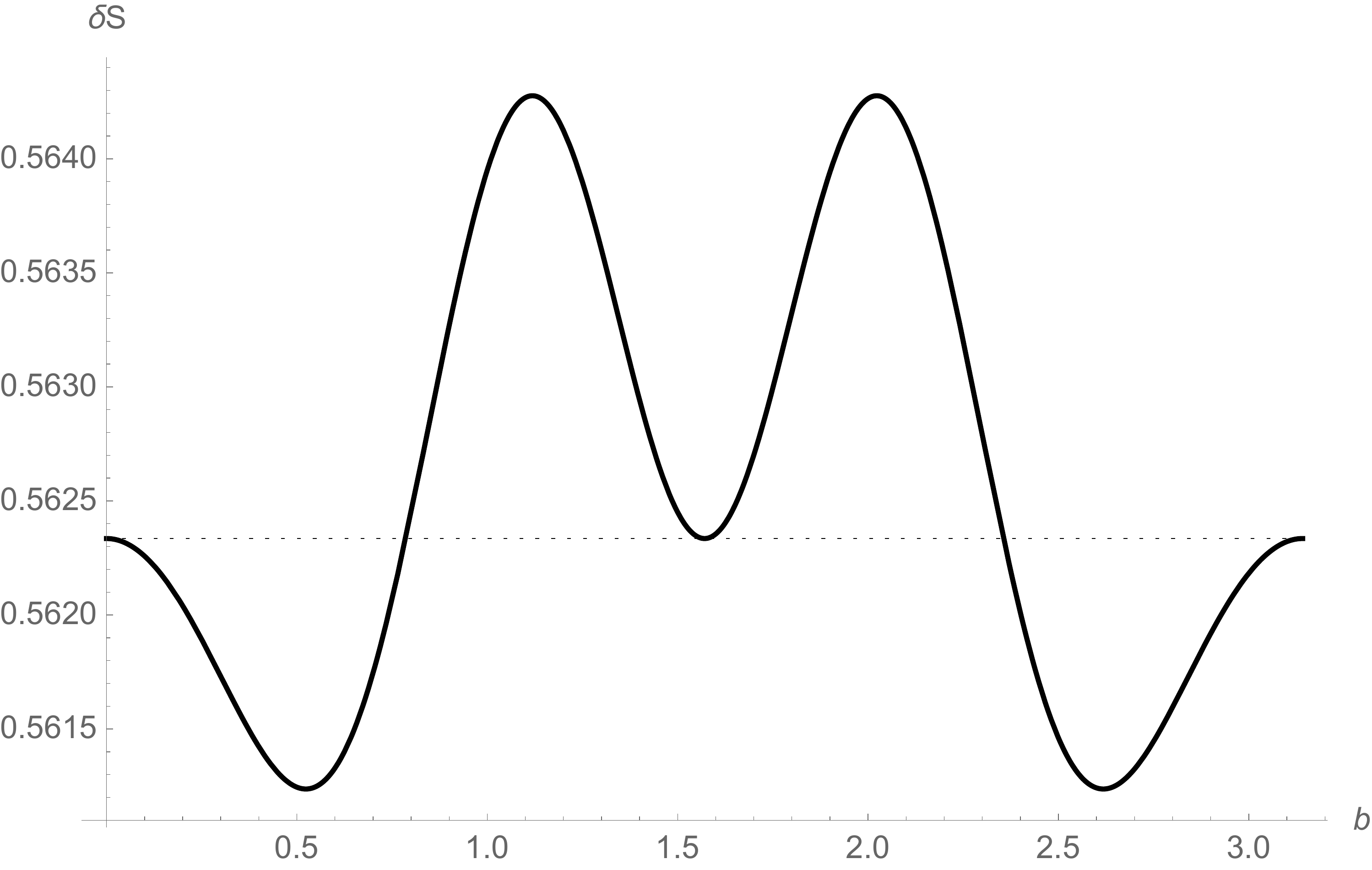}
\end{array}  
\)}
\caption{The entropy jump $\delta S$ as a function of $b$ in Planck units for $\gamma=\mu=\Delta_-=1$.
The small $b\ell_p$ behaviour in \eqref{bbb} is apparent. The entropy is periodic for $b\ell_p\in [0,\pi]$ as expected from \eqref{61}.}
\label{BBf}
\end{figure}

The general entropy jump for arbitrary (not necessarily small $\Lambda$) can be computed explicitly. Its value is bounded by $\log(2)$ in our model. Finally, the energy is conserved through the big bang and during all the dynamical evolution for arbitrary values of $b_0$. 
The decoherence and entanglement which can be interpreted as an information loss happens without energy spending as required by the 
scheme put forward in \cite{Perez:2014xca}.

\section{Discussion}

We have seen that one can precisely realize the scenario put forward in \cite{Perez:2014xca} for the resolution of Hawking's information loss paradox in quantum gravity in the context of loop quantum comology. The key feature making this possible is the existence of additional degrees of freedom with no macroscopic interpretation which unavoidably entangle with the macroscopic degrees of freedom during the dynamical evolution and lead to a reduce density matrix whose entropy grows. The fundamental description is unitary but the effective description---that does not take the microscopic degrees of freedom into account and hence is analogous to the QFT description of BH evaporation---evolves pure states into mixed states.
The microscopic degrees of freedom in the toy model are not introduced by hand, their existence is intimately related to the peculiar choice of representation of the fundamental phase space variables that leads to singularity resolution \cite{Bojowald:2001xe}. Moreover, such representation mimics the one used in the full theory of loop quantum gravity \cite{Lewandowski:2005jk} where also one expects such extra residual and microscopic degrees of freedom to exist and remain hidden to low energy coarse grained observers describing physics in terms of an effective QFT. 

From a more general perspective we expect this scenario to transcend the framework of loop quantum gravity: in any approach to quantum gravity, where spacetime geometry is emergent \footnote{For instance in the causal sets approach \cite{Bombelli:1987aa}, or in the context of Jacobson's ideas about emergence \cite{Jacobson:1995ab} (where, incidentally, in both cases unimodular gravity is the natural emergent structure), in causal dynamical triangulations \cite{Ambjorn:2004qm},  group field theory \cite{Oriti:2011jm}, etc.} from more fundamental discrete degrees of freedom, the effect (precisely illustrated here by our toy model) would generically occur.   

These results extrapolated to the context of black hole formation and evaporation suggest a simple  resolution of the information paradox that avoids the pathological features of other proposals. For instance  the possible development of  firewalls \cite{Almheiri:2012rt, Braunstein:2009my} or the risks of information cloning that holographic type of scenarios must deal with \cite{Marolf:2017jkr} are completely absent here. As decoherence in our model takes place without diffusion \cite{Unruh:2012vd}, the usual difficulties \cite{Banks:1983by} with energy conservation in the purification process are avoided along the lines of \cite{Unruh:1995gn, Unruh:2012vd}, yet in a concrete quantum gravity framework (hence without the problems faced by the QFT approach \cite{Hotta:2015yla, Wald:2019ygd}). 

We notice that the possibility of decoherence illustrated in the present model also suggest the possibility of diffusion into the underlying Planckian structure, such diffusion might have, in suitable situations, important consequences at large scales as argued in a series of recent papers \cite{Josset:2016vrq, Perez:2018wlo, Perez:2017krv}. The present model is very simplistic and realizes an example where such diffusion is not possible  due to (unimodular) energy conservation and the fact that the microscopic degrees of freedom do not contribute independently to the Hamiltonian. Nevertheless, one could generalize these models easily in order to include diffusion. This possibility is under current investigation and we plan to report the results elsewhere.

\section{Acknowledgments}

We thank P. Martin-Dussaud for finding for us the proof of useful inequalities in reference \cite{QE}. We thank M. Geiller, C. Rovelli, S. Speziale, M. Varadarajan, and E. Wilson-Ewing for stimulating discussions and Pietro Don\`a for the key remark  that the notion of coarse graining entropy that we had initially in mind could be written as standard entanglement entropy.

%
%
%

\appendix

\section{Entropy calculation}\label{entropia}

From the form of matrix \eqref{rere} we conclude that block diagonal with a single non trivial $2\times 2$ block of the form
\be\label{abus}
\rho_{\rm inR}=\mathbf{D}\otimes 0,
\ee
where 
\be
\mathbf{D}\equiv \frac{1}{2} \ket{\Psi_+(t)}\bra{\Psi_+(t)}+\frac{1}{2} \ket{\Psi_-(t)}\bra{\Psi_-(t)}.
\ee
One can compute the eigenvalues of the Hermitian matrix $\mathbf{D}$ (whose matrix elements can be found for instance via the Gram-Schmidt algorithm)  and find that they are given by $(1\pm |d|)/2$ with
\be
d\equiv \int db \, \overline \psi(b) \psi(b)  e^{i [E_{+}(b) - E_{-}(b)]t}.
\ee
Thus he entropy is
\be
S=-\frac{(1+|d|)}{2}\log\Big{[}\frac{(1+|d|)}{2}\Big{]}-\frac{(1-|d|)}{2}\log\Big{[}\frac{(1-|d|)}{2}\Big{]}.
\ee
The previous expression can be put in the form of \eqref{rentro}  via the substitution $\delta=1-|d|$ which is equation \eqref{pito}.
\section{Bound states and oscillating universes}\label{ultima}

Assuming $\mu >0$ and taking $s= -1$ then equation \eqref{shishi} has a solution with negative energy (a bound state) or a cosmological constant solution with energy
\be
 E_{s,\mu} (b)=-\frac{3 V_0}{8\pi G \gamma^2}  \frac{1}{ \Delta_s \ell_p^2} \left( {\sinh( \sqrt{\Delta_s} \ell_p b_{\mu} )} \right)^{2}
\ee 
with $b_\mu$ satisfying the following equation
\be
\frac{3}{8\pi \gamma^2}   \frac{\sinh(2 \sqrt{\Delta_s} b_\mu \ell_p)}{2 \sqrt{\Delta_s}}= {\mu} . 
\ee
The wave function for such solution is
\be
\Psi_{\rm BU}(\nu)= \sN {\exp(-|\nu| b_\mu)},
\ee
where $\sN$ is a normalization factor. Notice that the amplitude decays exponentially for large volumes. This is a universe of Planckian size to which the name baby universe would seem to apply.

\section{An alternative definition of coarse graining entropy}\label{cge}

There is a natural alternative notion of entropy associated the situation where the set of states that are seen as equivalent by the observer can be organised in `bins' mathematically defined by a set of projection operators $P_i$. From a given density matrix $\rho$ one defines the coarse-grained density matrix  \cite{PhysRevA.99.010101, PhysRevA.99.012103}\be\label{cge}
\rho^{\rm \va CG}\equiv \sum_i \frac{{\rm Tr}[P_i\rho]} {\rm Tr[P_i]} P_i,
\ee
which satisfies all the properties of a density matrix, and gives equal probability to all the elements in each given bin of equivalent states. It flattens the probability distribution in each set of equivalent states representing the complete ignorance of the coarse-grained observer concerning the difference of such states. The coarse-grained entropy is defined as
\be\label{cge-s} S(\rho^{\rm \va CG})\equiv-{\rm Tr}[\rho^{\rm \va CG} \log(\rho^{\rm \va CG})].\ee
It follows from standard results \cite{QE} that 
 \be\label{scg}
 S(\rho^{\rm \va CG})\ge S(\rho)\equiv -{\rm Tr}[ \rho \log( \rho)].
 \ee
This definition can of course be used in both the cases studied in Section \ref{sisi} and Section \ref{soso}. We do this now for completeness

\subsection{Coarse graining entropy in the situation of Section \ref{sisi}}
\begin{figure}[h] \centerline{\hspace{0.5cm} \(
\begin{array}{c}
\includegraphics[width=10cm]{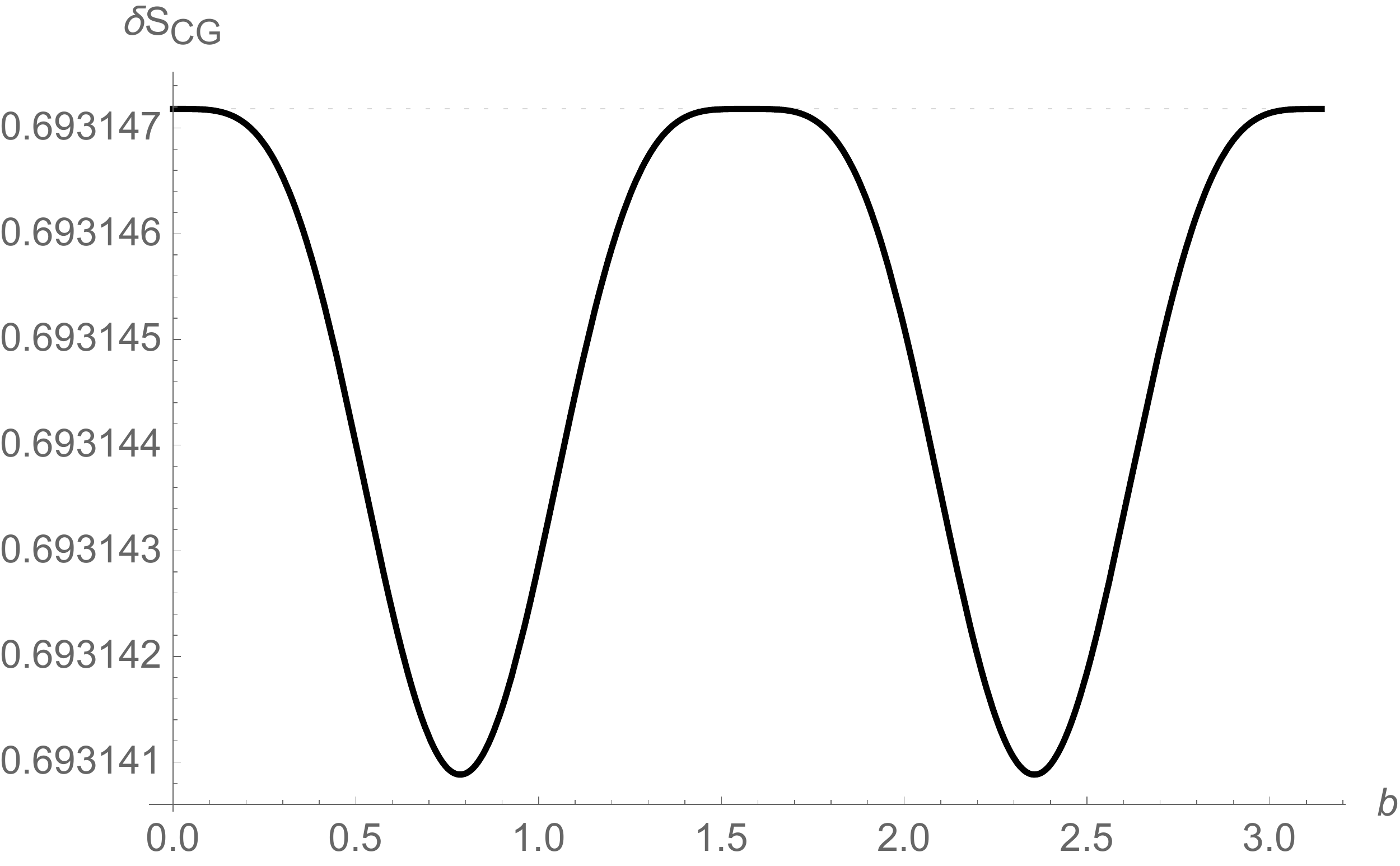}
\end{array}  
\)}
\caption{Coarse-graining entropy jump $\delta S_{\rm CG}$  due to matter coupling (modelled by our simple interaction) after the bing-bang as a function of $b\ell_p$  for $\gamma=\mu=\Delta_-=1$. The indistinguishability of the $\Gamma^1$ and $\Gamma^2$ components of the wave function for a coarse observer introduces an information loss that is quantified by the coarse graining entropy jump. The coarse graining entropy is constant $S^{in}_{\rm CG}={\rm log}(2)$ before and evolves to $S^{out}_{\rm CG}={\rm log}(2)+\delta S_{\rm CG}$.}
\label{BBCoarse}
\end{figure}

We want to define the coarse grained density matrix now. In our simple case we have only two projectors encoding the equivalence under our coarse-graining of $1$ and $2$ states. Namely, in terms of assymptotic states used to write the previous finite dimensional ($4\time 4$) matrix the projectors are
\be
P_1\equiv \ket{b_0, 1}\bra{b_0, 1}+\ket{b_0,2}\bra{ b_0, 2}=\left(\begin{matrix}1& 0 & 0 & 0\\  0& 0& 0& 0\\  0& 0& 1& 0\\  0& 0& 0& 0\end{matrix}\right),
\ee 
and
\be
P_2=\mathbf{1}-P_1\equiv \ket{-b_0,1}\bra{-b_0, 1}+\ket{-b_0,2}\bra{-b_0,2}=\left(\begin{matrix}0& 0 & 0 & 0\\  0& 1& 0& 0\\  0& 0& 0& 0\\  0& 0& 0& 1\end{matrix}\right).
\ee 
From the definition of the coarse grained density matrix \eqref{cge} we obtain
\ba
{\bm \rho}^{CG}_{in}=\left(
\begin{array}{cccc}
 \frac{1}{2} & 0 & 0 & 0 \\
 0 & 0 & 0 & 0 \\
 0& 0 & \frac{1}{2} & 0 \\
 0 & 0 & 0 & 0 \\
\end{array}
\right) \ \ \ \rightarrow \ \ \ {\bm \rho}^{CG}_{out}=\frac{1}{4}\left(\begin{matrix}  1+|B(b_0)|^2 & 0 &0  & 0 \\
 0 & |A(-b_0)|^2  & 0 & 0\\
 0 &  0 &1+|B(b_0)|^2   & 0 \\
 0& 0& 0& |A(-b_0)|^2 \end{matrix}\right).\n\ea
One can now straightforwardly compute the coarse graining entropy (\ref{cge-s}). We observe that it is constantly equal to $\log(2)$ before the big-bang and then it jumps across the {\em would-be-singularity} by an amount that depends on $b\ell_p$ when the expressions \eqref{eq152} and \eqref{61-s} are used. The result  is illustrated in the Figure \ref{BBCoarse}. To leading order in $\Lambda$ (or equivalently in $b\ell_p$) we have
\be
\delta S_{CG}=\log (2)-\frac{9 \Delta ^2
   \ell_p^4}{8192 \pi ^4 \gamma ^4 \mu ^4} \Lambda^2+\sO(\Lambda^2\ell_p^4)
\ee
The reason for this is the dynamical difference in the evolution of the $1$ and $2$ components of the wave function. It is important to point out that the entropy jump is generic: the qualitative result would remain the same if we had solved the more realistic model with the scalar field \eqref{scaly} or any other matter coupling. 

 \section{The free massless scalar field (numerical results)} \label{scalarf}
 
The case of gravity coupled to a scalar field \eqref{hf} can be solved numerically. The Hamiltonian is given by

\begin{equation}\label{num_matter_ham}
 \Hat{H} \psi(\nu)
    = \frac{3 V_0}{32 \pi G } \frac{1}{\gamma^2 \lambda^2}    \left[ \Psi(\nu - 4 \lambda) + \Psi(\nu + 4 \lambda)  - 2\Psi(\nu) \right] + \frac{V_0 p_{\phi}}{(2 \pi \ell_{Pl}^{2}  \gamma)^2 \nu^{2}} (1-\delta_{\nu 0})\Psi(\nu),
\end{equation}
where the factor $(1-\delta_{\nu 0})$ in potential is there to regularize the big-bang divergence at $\nu=0$ (see \cite{WilsonEwing:2012bx, Singh:2013ava}). Qualitative results remain the same for any of the regularization prescriptions discussed in the literature (in particular for those introduced in \eqref{77}).

As in the case of the ultra-local Hamiltonian \eqref{eqq36}, when the initial state is an eigenstate of the field momentum operator $p_{\phi}$, the problem reduces to a scattering problem. The setup of the problem is the same as in Section \ref{sca}: for simplicity we consider only states supported on the superposition of two lattices $\Gamma_{1}= \Gamma_\Delta^{\epsilon = 0}$ and $\Gamma_{2} =\Gamma_\Delta^{\epsilon=2\lambda}$. Initial states are Gaussian wavepackets shifted $2 \lambda $ one from the other. In other words, if $\phi(\nu,\Gamma_{1})$ , $\phi(\nu,\Gamma_{2})$ are Gaussian wavepackets with support on the lattices $\phi(\nu,\Gamma_{1})$ and $\phi(\nu,\Gamma_{2})$ respectively we have that $\exp(i \lambda b)\vartriangleright \phi(\nu,\Gamma_{1}) = \phi(\nu,\Gamma_{2})$. $\phi(\nu,\Gamma_{1})$ is peaked around $\nu_{0}$ while $\phi(\nu,\Gamma_{2})$ is peaked around $\nu_{0} - 2 \lambda$.

\begin{figure}[h] \centerline{\hspace{0.5cm} \(
\begin{array}{c}
\includegraphics[width=8cm]{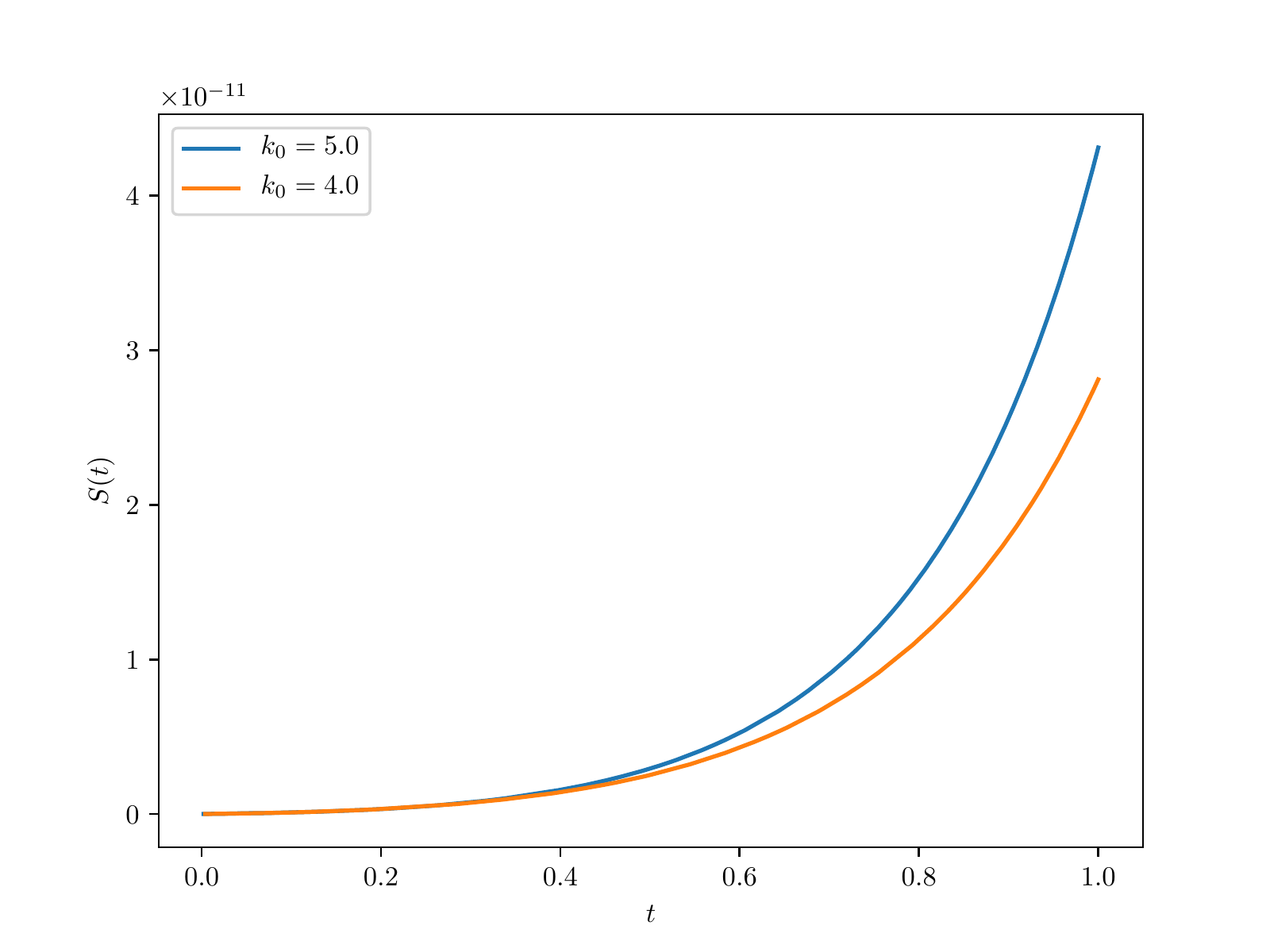} \end{array}\ \ \ \ \begin{array}{c}
 \includegraphics[width=8cm]{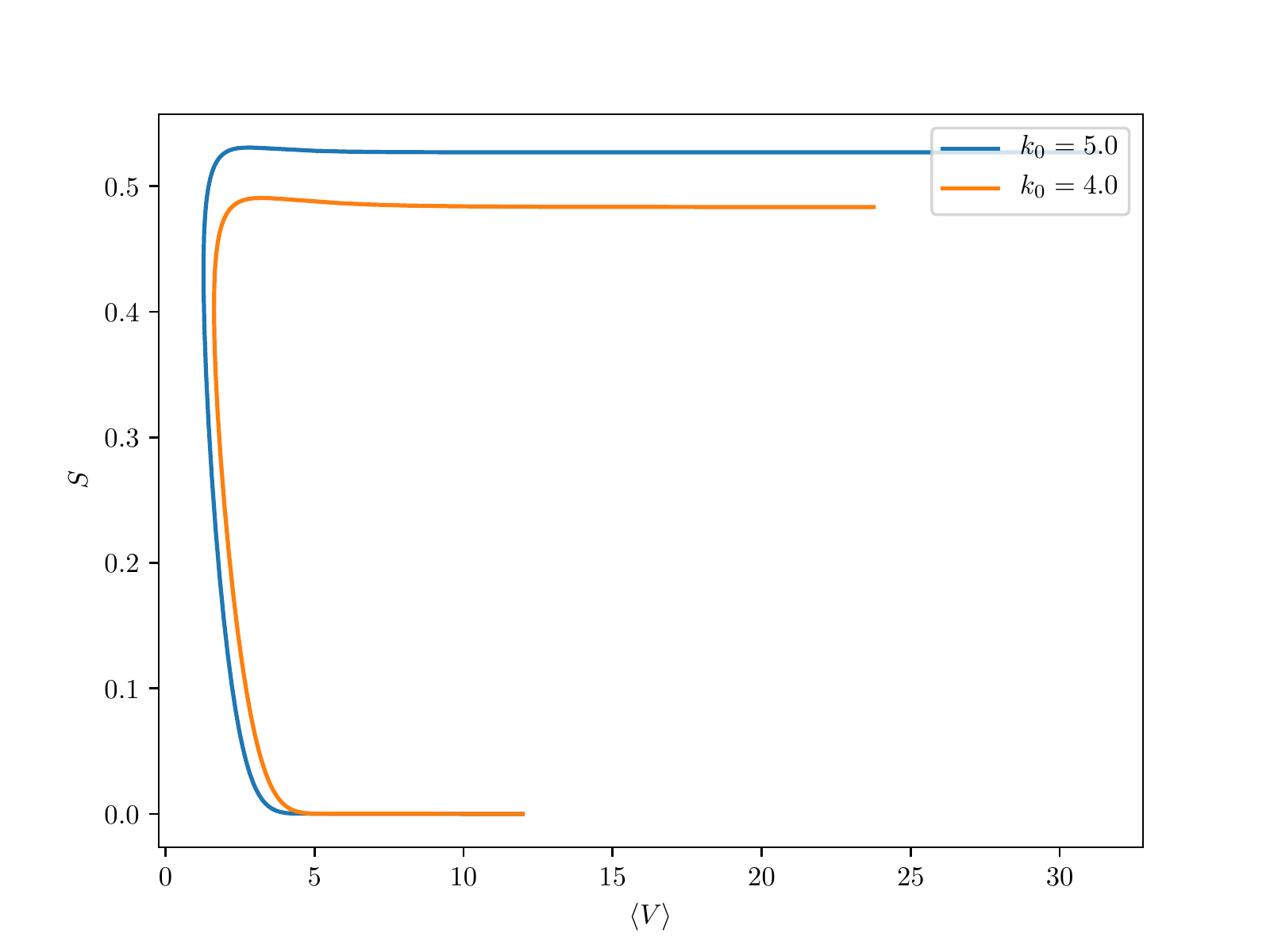}
\end{array}
\)}
  \caption{ \textbf{Left Panel:} Entropy $S$ with respect to unimodular time $t$ away from the bounce, where the matter interaction is almost negligible. For states with higher cosmological constant entropy grows faster. \textbf{Right Panel:} Entropy $S$ with respect to the expected value of the volume $\langle V \rangle$ through the bounce. The entropy jump around the bounce is generic while the quantitative details depends on the initial state and the value $p_{\phi}$. For both cases $\gamma = 1, \ V_0=1, \ \nu_{0}=12, \ p_{\phi}=2, \lambda = {\sqrt{2}}/{40}$ in natural units.}
\label{campo_escalar_entropy}
\end{figure}

We see that entropy grows gradually as the Universe approaches the bounce where the matter Hamiltonian starts changing abruptly from one lattice site to another. Thus, components of the quantum state supported on different lattices will 'see' different potentials and entropy will grow rapidly.
This effect depends on the semiclassical curvature of the initial state. Initially, when the Universe is sufficiently large and for low cosmological constant $\Lambda$, decoherence will become important closer to the bounce than for states with higher $\Lambda$. The wavefunction of states with higher cosmological constant oscillate faster thus amplifying differences between the matter Hamiltonian when proved by two lattices with different $\epsilon$. This behavior is illustrated in Fig \ref{campo_escalar_entropy}.

\end{document}